\documentclass{aa}  
\usepackage{natbib}
\bibpunct{(}{)}{;}{a}{}{,}
\usepackage{graphicx}
\usepackage{enumitem}
\usepackage[percent]{overpic}
\usepackage{epstopdf}

\begin{document}
\authorrunning{Chatzistergos et al.}
\titlerunning{Photometric calibration and CLV compensation of Ca~II~K observations.}
\title{Analysis of full disc Ca~II~K spectroheliograms. \\I. Photometric calibration and CLV compensation}
\author{Theodosios Chatzistergos\inst{1}, Ilaria Ermolli\inst{2}, Sami K. Solanki\inst{1,3},  Natalie A. Krivova\inst{1}}
\offprints{Theodosios Chatzistergos  \email{chatzistergos@mps.mpg.de}}
\institute{Max-Planck-Institut f\"{u}r Sonnensystemforschung, Justus-von-Liebig-Weg 3,	37077 G\"{o}ttingen, Germany \and INAF Osservatorio Astronomico di Roma, Via Frascati 33, 00078 Monte Porzio Catone, Italy \and School of Space Research, Kyung Hee University, Yongin, Gyeonggi 446-701, Republic of Korea}
\date{}

\abstract
{Historical Ca~II~K spectroheliograms (SHG) are unique in representing long-term variations of the solar chromospheric magnetic field. They usually suffer from numerous problems and lack photometric calibration. Thus accurate processing of these data is required to get  meaningful results from their analysis.}
{In this paper we present an automatic processing and photometric calibration method that provides precise and consistent results when applied to historical SHG. }
{ The proposed method is based on the assumption that the centre-to-limb variation of the intensity in quiet Sun regions does not vary with time. We tested the accuracy of the proposed method on various sets of synthetic images that mimic   problems encountered in historical observations. 
We also tested our approach  on a large sample of images randomly extracted from seven different SHG archives.} 
{The tests carried out on the synthetic data show that the maximum relative errors of the method are generally $<6.5$\%, while the average error is $<1$\%, even if rather poor quality observations are considered. In the absence of strong artefacts the method returns images that differ from the ideal ones by $<2$\% in any pixel. The method gives consistent values for both plage and network areas.
We also show  that our method returns consistent results for images from different SHG archives. 
} 
{Our tests show that the proposed method is more accurate than other methods presented in the literature. 
Our method can also be applied to process images from photographic archives of solar observations at other wavelengths than Ca~II~K. 
}
\maketitle

%%%%%%%%%%%%%%%%%%%%%%%%%%%%%%%%%%%%%%%%%%%%%%%%%%%%%%%%%%%%%%%%%%%%%%%%%%%%
%%%%%%%%%%%%%%%%%%%%%%%%%%%%%%%%%%%%%%%%%%%%%%%%%%%%%%%%%%%%%%%%%%%%%%%%%%%%
\section{Introduction}
\label{sec:intro}
The Sun's magnetic field is concentrated into strong-field elements, or flux tubes, in addition to a weaker, more turbulent background.
The strong-field component is thought to be mainly responsible for solar activity. In the lower solar atmosphere, this strong-field flux manifests itself as dark features, such as sunspots and pores, as well as bright structures forming faculae (or plage) and the network. With modern magnetographs and spectropolarimeters both components can be reliably observed. However, before Kitt Peak National Observatory started recording magnetograms in 1974, typically only sunspot records are available. This has negative repercussions for a number of research fields. One of these is the solar influence on climate through its variable irradiance \citep{haigh_sun_2007,solanki_solar_2013,ermolli_recent_2013}.

Regular space-based measurements have shown that the total solar irradiance (TSI, the spectrally integrated solar radiative flux at the top of the Earth's atmosphere at the mean  distance of one astronomical unit)  varies on different time scales, from minutes to decades \citep{frohlich_total_2013,kopp_magnitudes_2016}.
The variability at time scales greater than a day is attributed to changes in the surface distribution of the solar magnetic field.
Models have been developed to reconstruct the solar irradiance by accounting for the competing contributions from dark and bright magnetic features on the solar surface. These models have been successful in reproducing the measured TSI \citep{krivova_reconstruction_2003,ermolli_recent_2011,ball_reconstruction_2012,chapman_modeling_2013,yeo_reconstruction_2014}. 
However, full-disc images of the Sun, typically used by such models, are only available for the last few decades. Reconstructions of past irradiance changes usually rely on observations of sunspots \citep{lean_reconstruction_1995,solanki_reconstruction_2000,wang_modeling_2005,krivova_reconstruction_2007,krivova_reconstruction_2010,dasi-espuig_modelling_2014,dasi-espuig_reconstruction_2016}, or concentrations of cosmogenic radionuclides, e.g. $^{14}$C and $^{10}$Be, \citep{steinhilber_total_2009,delaygue_antarctic_2011,shapiro_new_2011,vieira_evolution_2011}. 
However, such data provide information about the bright regions only indirectly and on the basis of strong assumptions.

The time-series of available bright magnetic feature data can be potentially extended with the help of daily full-disc Ca~II~K spectroheliograms (SHG, hereafter) that have been stored in photographic archives at various sites since the beginning of the 20th century.
Indeed, owing to the relation  between the Ca~II~K brightness and the magnetic field strength averaged over a pixel \citep[e.g.][]{babcock_suns_1955,skumanich_statistical_1975,schrijver_relations_1989,harvey_magnetic_1999,loukitcheva_relationship_2009,kahil_brightness_2017}, the SHG can be used to describe the temporal and spatial evolution of the solar surface magnetic field, and specifically its bright component as seen in the chromosphere.
Since SHG allow direct measurements of the surface coverage and, with a proper calibration, also the photometric properties of bright magnetic features, the historical photographic archives are an important resource for studies of solar magnetism over the last century and the activity and irradiance variations it produced.
This has led to an increasing interest towards the recovery and preservation of  these early  observations, which has recently prompted the digitization of  various SHG archives, such as those at the Arcetri \citep{ermolli_digitized_2009}, Coimbra \citep{garcia_synoptic_2011}, Kodaikanal \citep{priyal_long_2013}, Meudon \citep{mein_spectroheliograms_1990}, Mitaka \citep{hanaoka_long-term_2013}, and Mt Wilson \citep{bertello_mount_2010} observatories.

\begin{figure}[h!]
	\centering
	\includegraphics[width=1\linewidth]{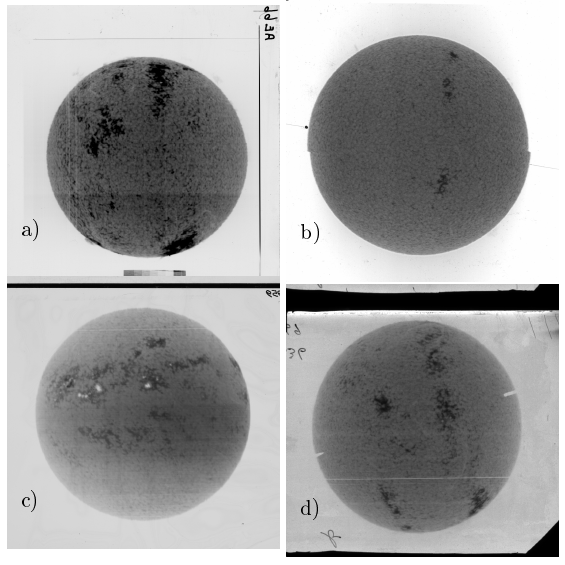}
	\caption{Examples of digitised SHG taken at the: (\textbf{a)}) Arcetri (06/09/1957), (\textbf{b)}) Kodaikanal (14/09/1912), (\textbf{c)}) Mitaka (25/01/1959), and (\textbf{d)}) Mt Wilson (01/01/1969) observatories.}
	\label{fig:01_01_shg_examples}
\end{figure}

Figure \ref{fig:01_01_shg_examples} presents typical examples of the images derived from the digitisation of SHG (hereafter referred to as digitised images). 
These images show  non-uniformity across the field-of-view of solar and non-solar origin, as well as (non-solar) spots and scratches on the photographic support.

\begin{figure}
	\centering
	\includegraphics[width=1\linewidth]{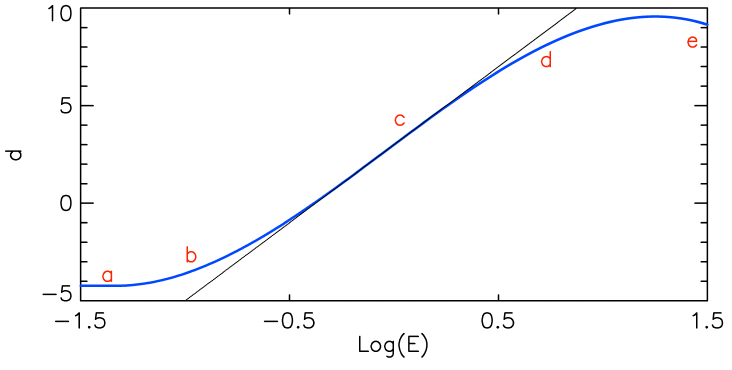}
	\caption{Typical form of a characteristic curve (blue). The labels on the curve denote the different exposure regions: (a) fog level, (b) under-exposure, (c) proper exposure, (d) over-exposure, (e) solarisation. The black line has the same slope as the region of proper exposure. }
	\label{fig:cc}
\end{figure}

Analysis of SHG requires accounting for these  non-uniformities and for the non-linear relation between the  blackening of the photographic  
support (either glass plate or film, hereafter referred to as plate) and the flux of incident solar radiation during the plate's  exposure. 
This relation is expressed by the characteristic curve (or Hurter and Driffield curve, CC hereafter), which is specific of each photographic observation \citep{dainty_image_1974} as it depends on a variety of factors, e.g. the exposure time, the gelatine of the photographic plate, the composition of the  developers (reducing and fixing baths), the duration of plate development and the temperature, and degree of stirring during this step. The CC has in general a sigmoid shape. A typical example of a CC is shown in Fig.~\ref{fig:cc} \citep[see e.g. Sect. 3 or][for more details]{dainty_image_1974}.

Until recent times, the CC was mainly derived using specific exposures (step-wedges, hereafter) acquired with known relative intensity ratios, in addition to the  archived observation \citep[e.g.][]{fredga_comparison_1971,kariyappa_variability_1994,kariyappa_contribution_1996,worden_evolution_1998,giorgi_calibration_2005,ermolli_digitized_2009}. 
However, in most cases acquisition of these exposures started many years after the start of solar observations. 
For example, synoptic SHG were taken regularly since  1915 at the Mt Wilson Observatory, but calibration exposures were imprinted on the plates only from 1961 onwards; at the Kodaikanal site observations started in late 1904 and step-wedges measurements in 1958. 
Therefore, the bulk of these data lacks specific information on the photometric calibration. 

Nevertheless, there have been attempts to calibrate
 the data lacking wedges.
\citet{priyal_long_2013} calibrated Kodaikanal data by applying an average CC derived from the available step-wedge exposures on all images. 
\citet{ermolli_comparison_2009} applied the method presented by \citet{mickaelian_digitized_2007} for calibration of photographic plates of star surveys, which is based on information stored in the unexposed and darker parts of the plate. 
\citet{tlatov_new_2009} suggested to calibrate SHG by using a linear relation between a standardized centre-to-limb variation (CLV, hereafter) profile published by \cite{pierce_solar_1977} (derived with a 2nd degree polynomial fit and corresponding to 390.928 nm) and the values on the analysed image at two positions. 

Various methods have also been used to process Ca~II~K images and compensate them for the CLV.
Some examples are computations that result in radially symmetric backgrounds \citep{brandt_determination_1998,walton_processing_1998,johannesson_10-year_1998,denker_synoptic_1999,zharkova_full-disk_2003}, application of a median filter \citep{lefebvre_solar_2005,bertello_mount_2010,chatterjee_butterfly_2016}, 2D polynomial fittings to the entire image \citep{worden_plage_1998,caccin_variations_1998}, or interpolation of the mode pixel values within radial and azimuthal disc sectors \citep{tlatov_new_2009}.
However, all of these methods are unable to thoroughly account for the artefacts affecting historical data.
\citet{worden_evolution_1998} presented a method that is able to account for the inhomogeneities in the images by fitting a 1D 5th degree polynomial along image rows and columns to density values of the original observation and on the image resulting from its  $45^\circ$ rotation. Regions with values outside the  $\pm 2\sigma$  interval of the image, where $\sigma$ is the standard deviation, were excluded from the analysis and the background is estimated by applying a median, low-pass filter to the average of the fit. 
Other studies applied variations and combinations of methods previously described in the literature \citep[e.g.][]{ermolli_comparison_2009,singh_determination_2012,priyal_long_2013}.
For instance, \citet{priyal_long_2013} used the method of \cite{denker_synoptic_1999} and a modified version of the approach suggested by \citet{worden_evolution_1998}, applying a 3rd degree polynomial fit instead of the originally proposed 5th degree polynomial.

The long list of references above show that many methods have been employed for processing and calibrating historical Ca~II~K observations. However, their aptness and accuracy has generally been discussed only qualitatively.

In this paper we present a new method to calibrate the historical photographic Ca~II~K SHG as well as a novel method to compensate for various patterns on the data.  These patterns include the CLV of the solar intensity and patterns of non-solar origin introduced by observational and archival processes. 
This  method, which  works without  knowledge of the specific CC of the analysed plate and in the absence of step-wedge imprints,  is based on the computation of the CLV of density values on quiet Sun (QS, hereafter) regions. Importantly, we rigorously test our technique against modern CCD-based data artificially degraded to correspond to photographically obtained SHG.

The historical and modern observations used in this study are described in Sect. \ref{sec:data}.
The method is presented in Sect. \ref{sec:method}.
We assess the accuracy of the method with synthetic data that have known characteristics and artefacts and test the proposed method on a large sample of SHG in Sect. \ref{sec:results}. We compare our method with other approaches presented in the literature in Sect. \ref{sec:comparisonwithothermethods} and draw our conclusions in Sect. \ref{sec:conclusions}.

\newcounter{tableid}
\section{Observational data}
\label{sec:data}
\begin{table*}[th!]
\centering
	\caption{List of Ca~II~K digitised image archives considered in our study.}
	\label{tab:observatories}
	\begin{tabular}{l*{7}{c}}
		\hline\hline
		Observatory & Years & Number of images & Spectral width  & Disc size& Data type & Pixel scale  &Reference\\
		&   & & [nm] &  [mm]& [bit]& [$"/$pixel] \\%&ermolli2009\\
		\hline
		Arcetri 				& $1931-1974$ & 5141  & 0.03 & $ 65$	& 16 & 2.4    & \addtocounter{tableid}{1} \thetableid\\
		Coimbra 				& $1994-1996$ & 79 	& 0.016&  $87$	& 16 & 1.8    & \addtocounter{tableid}{1} \thetableid\\ 
		Kodaikanal\tablefootmark{a}  & $1907-1999$ & 22164 & 0.05 & $ 60$	& 8  & 1.3 	  & \addtocounter{tableid}{1} \thetableid\\ 
		Meudon 					& $1980-2002$ & 5727 	& 0.015& 86 	& 16 & 1.5 	  & \addtocounter{tableid}{1} \thetableid\\ 
		Mitaka 					& $1917-1974$ & 8585  & 0.05 &   variable		&16, 8\tablefootmark{b}&0.9, 2.9\tablefootmark{b} & \addtocounter{tableid}{1} \thetableid\\ 
		McMath-Hulbert 					& $1942-1979$ & 5593 & 0.01 & 17.3 & 8 & 3 & \addtocounter{tableid}{1} \thetableid \\ 
		Mt Wilson 				& $1915-1985$ & 36147 & 0.02 & $50$	& 16 & 2.7	  & \addtocounter{tableid}{1} \thetableid\\
		Rome/PSPT 				& $2000-2014$ & 2000  & 0.25 &$\sim27$& 16 & 2\tablefootmark{c} 	  & \addtocounter{tableid}{1} \thetableid\\
		\hline
	\end{tabular}
	\tablefoot{Columns are: name of the observatory, period of observations, number of images, spectral width of the spectrograph/filter, the size of the solar disc on the plate, digitisation depth, pixel scale of the digitised file, and the bibliography entry.
		\tablefoottext{a}{These data have more recently been re-digitised with 16-bit data type extending the dataset from 1904 to 2007 \citep{priyal_long_2013,chatterjee_butterfly_2016}.}
	\tablefoottext{b}{The two values correspond to the period before and after 02/03/1960, when the images were stored on photographic plates and on 24x35 mm sheet film, respectively.}
	\tablefoottext{c}{The pixel scale is for the resized images.}}
	\tablebib{\addtocounter{tableid}{-8}\addtocounter{tableid}{1} (\thetableid)~\citet{ermolli_digitized_2009}; \addtocounter{tableid}{1} (\thetableid) \citet{garcia_synoptic_2011}; \addtocounter{tableid}{1} (\thetableid) \citet{makarov_secular_2004}; \addtocounter{tableid}{1} (\thetableid) \citet{mein_spectroheliograms_1990}; 
		\addtocounter{tableid}{1} (\thetableid) \citet{hanaoka_long-term_2013}; \addtocounter{tableid}{1} (\thetableid) \citet{fredga_comparison_1971}; \addtocounter{tableid}{1} (\thetableid) \citet{bertello_mount_2010}; \addtocounter{tableid}{1} (\thetableid) \citet{ermolli_photometric_2007}.
	}
\end{table*}

The historical observations, on which our technique is tested, are the digitised SHG from the Arcetri (Ar), Coimbra (Co), Kodaikanal (Ko), Meudon (Me), Mitaka (Mi), McMath-Hulbert (MM), and Mt Wilson (MW)  observatories. 
These images were taken in the Ca~II~K  line at $\lambda$ = 393.367 nm, with nominal bandwidths ranging from 0.01 nm to 0.05 nm and scanning time of several minutes needed to cover the solar disc. 
The data, which were taken between February 1907 and May 2002 with different instrumental set-ups and stored mainly on photographic plates, were digitized by using various devices and methods. This results in solar images  with different sizes and characteristics depending on the archive.

The modern observations used to test our technique were taken at the INAF Osservatorio Astronomico di Roma (OAR) with the Precision Solar Photometric Telescope 
\citep[Rome/PSPT,][]{ ermolli_prototype_1998,ermolli_photometric_2007}. 
These data were acquired from August 2000 to October 2014,  by using an interference filter centred at the Ca~II~K line with a 0.25 nm bandwidth and the exposure time much shorter than for the historical observations, usually about  0.06 s.
We analysed images after the usual reduction steps (dark current removal and flat fielding) that greatly reduce instrumental effects and  after they were resized to half their original dimension (2 arcsec per pixel after resizing), to roughly match the size of the historical data. 

Key information about the datasets used in this study is summarised in Table \ref{tab:observatories}.

The available data include estimates of the solar radius $R$ and centre coordinates in the images. 
To avoid our results being affected by any uncertainties in the radius estimates, we only analysed the disc pixels within $0.99R$ ($0.98R$ for Ko). This corresponds to $\mu=\cos{\theta}=0.14$ ($\mu=0.2$ for Ko), where $\theta$ is the heliocentric angle.

\section{Method description}

\label{sec:method}
\label{sec:preprocessing}
The digitized SHG reflect the blackening degree of the emulsion, which is proportional to the transparency $T$ of the plate and is related to the density $d$ of the photographic emulsion since: 
\begin{equation}
\label{eq:densitytransparency}
d=\log{\left(1/T\right)}.
\end{equation}
The modern data represent the flux of incident radiation during the observation. 
The CC is defined as the relation $d = f (\log{E})$ (shown in Fig. \ref{fig:cc}), where $E=I\times t$ is the plate exposure, $I$ the incident energy per unit area and $t$ the exposure time. 
Assuming that the plate was exposed evenly, the CC becomes $d=f(\log{I})$. 
Therefore the photometric calibration of the data means applying on them the specific and appropriate relation between $d$ and $I$.

Our method of calibration and processing of the SHG is based on the assumption that, in these observations, the CLV of intensity in the internetwork regions, i.e. the quietest part of the QS, does not vary with time, in agreement with  
\citet{white_solar_1978,white_solar_1981}, \citet{livingston_suns_2003}, and \citet{livingston_sun-as--star_2007}. 
This assumption is also supported by the results of \citet{buhler_quiet_2013} and \citet{lites_solar_2014} that the internetwork magnetic flux, which is the main component of magnetic field populating these quietest parts of the solar surface, 
remains unchanged over the solar cycle.
Before applying the proposed method to the historical images, we convert their values to densities according to Eq. \ref{eq:densitytransparency}. 
Figure \ref{fig:nsb_steps}\textbf{a)} shows the density image corresponding to the Ko observation displayed in Fig. \ref{fig:01_01_shg_examples}\textbf{b)}.

The main steps of our method can be summarised as:
\begin{itemize}
	\item Calculation of the 2D map of QS regions in each image (including CLV and inhomogeneities);
	\item Extraction of the 1D QS CLV profile in each image;
	\item Construction of CC by relating the 1D QS CLV to a reference 1D radial intensity QS CLV from CCD-based Rome/PSPT observations;
	\item Calibration of the image using the CC;
	\item Compensation for the intensity CLV.
\end{itemize}

Visual inspection of historical data shows QS density patterns (2D QS background, hereafter)  that are in general strongly non-symmetric and inhomogeneous (see, e.g., Fig. \ref{fig:01_01_shg_examples}). 
This is due to a plethora of problems affecting the data, introduced either during the observation (e.g. partial coverage by clouds, vignetting, uneven movement of the slit),
the development (e.g. non homogeneous bathing, touching the plate before the process was finished), the storage period (e.g. dust accumulation, scratches, humidity, ageing burns), or the digitization (e.g.  dust, hair, loss of dynamic range) of the plate.
Since results from modern observations indicate the radial uniformity of radiative emission of the internetwork \citep[1D radial QS, ][]{livingston_limits_2008,peck_photometric_2015}, we assume that all departures of the 2D internetwork map in SHG from a radially symmetric map are  image artefacts of non-solar origin.

\begin{figure}
	\centering
	\includegraphics[width=0.7\linewidth]{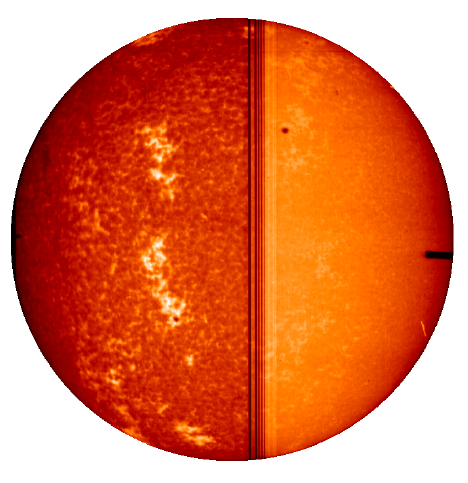}
	\caption{MW observation taken on 28/06/1979, where the right part of the image was taken with a broader bandwidth than the left part.}
	\label{fig:varbandwidthexample}
\end{figure}

It is worth mentioning that the bandwidths listed in Table \ref{tab:observatories} are nominal, while the real values can vary even within a single SHG.
The varying bandwidths mean that also the heights of the atmosphere that are sampled change. With broader bandwidths, the photospheric contribution is higher and plage regions appear less bright, while sunspots are more visible.
As an example, Fig. \ref{fig:varbandwidthexample} shows a MW observation where the right part of the image was taken with a broader bandwidth, either intentionally or due to instrument problems. If the bandwidth of the observations was varied or if they were centred at a slightly different wavelength this would have resulted in a different limb darkening profile.
However, as information on the real bandwidth is not available, in our analysis we have to assume that the bandwidth of the observations has remained constant over the disc and for all data within an archive.

%%%%%%%%%%%%%%%%%%%%%%%%%%%%%%%%

\subsection{Background computation}

\begin{figure*}[th!]
	\centering
	\includegraphics[width=1\linewidth]{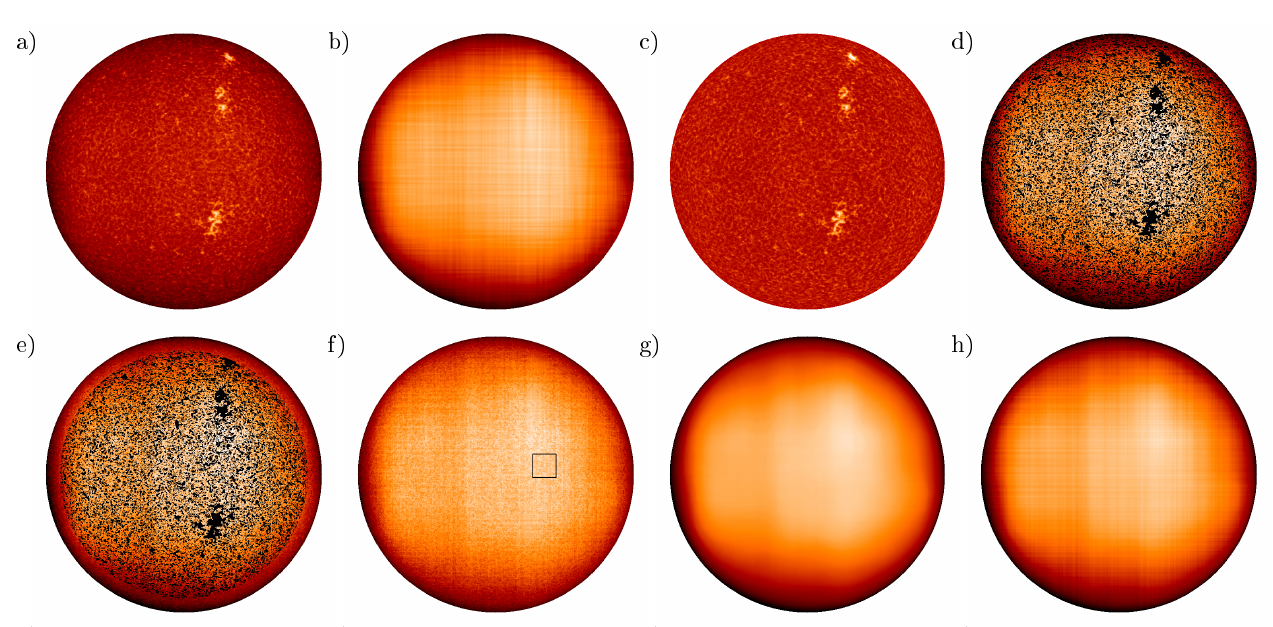}
	\caption{Selected processing steps of Ko observation taken on 14/09/1912: (\textbf{a)}) original density image, (\textbf{b)}) first estimate of background map (after \textbf{\textit{Step 1.4}}), (\textbf{c)}-\textbf{h)}) results of each step of the iterative process (\textit{Steps 2.1--2.5}), (\textbf{e)} and \textbf{f)}) correspond to \textit{Step 2.3}. The black square in panel \textbf{f)} shows the dimensions of the window used for the median filter in \textit{Step 2.4} (see Sect. \ref{sec:backgroundcomputation} for details).}
	\label{fig:nsb_steps}
\end{figure*}

\label{sec:backgroundcomputation}
We first derive the 2D QS background. For this, it is essential to accurately exclude active regions (AR, hereafter), otherwise there is a risk of overestimating the background due to the contribution of remaining AR.
The process of calculating the 2D QS  background can be outlined as follows:
\begin{itemize}
	\item Get a rough estimate of the background;
	\item Use the last estimate of the background to identify and exclude AR;
	\item Iteratively repeat the process of calculating the background and identifying the AR until sufficient accuracy is achieved;
\end{itemize}

The first rough estimate of the background is obtained in 3 steps.

\vspace{5pt}
\noindent {\it -- Step 1.1. } The solar disc is divided into azimuthal slices of $30^\circ$ that cover the disc in steps of $5^\circ$. Within each slice we apply a 5th degree polynomial fit of the form $d=f(\mu)$. 
The best fit values of $d$ are assigned to all pixels within a given slice. Results obtained from contiguous slices are gradually averaged and stitched together (this method will be referred to as rotating slices). 

\vspace{5pt} 
\noindent {\it -- Step 1.2. } We compute the density contrast image  $C^d_i = (d_i-d^{\mathrm{QS}}_i)/d^{\mathrm{QS}}_i$, where $d_i$ is the original density image and $d_i^{\mathrm{QS}}$ the density background resulting from {\it Step 1.1} at the $i$th pixel. 

\vspace{5pt}
\noindent {\it -- Step 1.3. }
We identify the regions in the density image lying outside 1$\sigma$ of the result of \textit{Step 1.2}. These are tentatively identified as AR and are excluded from the further analysis (note that the removal of non-AR pixels in this process does not influence the results; it is more important to discard as many AR pixels as possible).

\vspace{5pt}
\noindent {\it -- Step 1.4. } We apply a 5th degree polynomial fit to the density image, excluding AR defined in \textit{Step 1.3}, along each column and row of the image separately \citep[similarly to][but without the $45^\circ$ rotation of the disc]{worden_evolution_1998}. 
To all pixels of each analysed column/row we assign the density values resulting from the best fit.
We get a background map by stitching together the results obtained from the best fit; multiple values derived for the same location on the solar disc are averaged (we will refer to this method as column/row fittings). 

The calculations described in \textit{Steps 1.1--1.4} provide a first, rough estimate of the background. However, the identification of AR at this stage is rather inaccurate and the obtained values of the background around AR are overestimated. The varying contrast for different $\mu$ positions renders the identification near the limb less accurate.

Therefore in order to improve the AR identification and the calculation of the background, we iteratively repeat the following steps:

\vspace{5pt}
\noindent {\it -- Step 2.1. } 
We compute the density contrast image  $C^d_i = (d_i-d^{\mathrm{QS}}_i)/d^{\mathrm{QS}}_i$, where $d_i$ is the original density image and $d_i^{\mathrm{QS}}$ the density background resulting from the previous calculation at the $i$th pixel. During the first iteration $d^{\mathrm{QS}}$ is the rough background estimate (from \textit{Step 1.4}), afterwards we use the result of the previous iteration. 

\vspace{5pt}
\noindent {\it -- Step 2.2. } 
We remove the AR in the original density image, retaining only the pixels that fulfil the threshold criteria $C^d_i> \langle C^d \rangle {+l_1}\sigma$  or $C^d_i< \langle C^d \rangle {-l_2}\sigma$, where $l_1$,$l_2$ are the applied thresholds. 
For the first 3 iterations $l_1=1$ and $l_2=4$, afterwards $l_1=0.5$ and $l_2=1$. These threshold values were chosen following a series of tests which showed that they effectively removed plage, network and sunspots, leaving just internetwork regions.
See discussion in Sect. \ref{sec:NSB calculation}.

\vspace{5pt}
\noindent {\it -- Step 2.3. } 
To fill in the gaps left by the AR over the whole disc in the original density image we apply the column/row fittings on the image obtained in \textit{Step 2.2}. 
To avoid artefacts of the fit at the limb due to missing values, the gaps by the AR in the outer $0.1R$ are first filled in with the rotating slices method.
The column/row fittings method is repeated two more times to the residual of the density image to the calculated background in order to improve the accuracy of the computation (see discussion in Sect. \ref{sec:NSB calculation}). All three computations are summed together.

\vspace{5pt}
\noindent {\it -- Step 2.4. } 
We apply a  running window median filter on the image resulting from  \textit{Step 2.3}. The filter window width used is $R/6$ (shown with a rectangle in Fig. \ref{fig:nsb_steps}f; see discussion in Sect. \ref{sec:NSB calculation}). 
To avoid inconsistencies, the outer $R/12$ part of the disc is re-sampled outside the disc to fill the space between $R$ and $R+R/12$ and the pixel values of the re-sampled section are adjusted so that the median filter is not affected by the pixel values outside the disc.

\vspace{5pt}
\noindent {\it -- Step 2.5.} 
We identify dark and bright linear artefacts affecting many of the SHG (for instance see Fig. \ref{fig:01_01_shg_examples}\textbf{d)}) by separately fitting a polynomial to every row and column of the residual image between the original density image and the result of \textit{Step 2.4}, excluding the AR.  The fit is repeated 3 times as in \textit{Step 2.3}.

The sum of the maps derived in \textit{Steps 2.4} and \textit{2.5} is the final background map of each iteration.
The five-step processing described above is repeated until the difference between two subsequent background maps does not improve the accuracy of the QS background further. Usually 4 iterations allow lowering the relative unsigned mean difference between maps from two consecutive iterations to $<0.1$\%.

The result of the processing described so far is usually an asymmetric map (non-symmetric background, NSB hereafter) that describes the 2D QS background of the original image. 
The asymmetry is caused by a residual pattern that includes small- and large- scale inhomogeneities due to image problems, e.g. dark and bright bands and linear artefacts, stray light features, image gradients. Figure \ref{fig:nsb_steps} shows different steps of the processing on a randomly selected Ko SHG. In particular, the original density image is shown in panel \textbf{a)}, the result of the first column/row fittings process in panel \textbf{b)}, while panels \textbf{c)}--\textbf{h)} show the results of each step (2.1--2.5) of the iterative process. 

Finally, we apply the method of \cite{nesme-ribes_fractal_1996} on the residual image between the original and the NSB to identify and compensate any offset in the average level of the computed NSB and the analysed image. 
The method by \cite{nesme-ribes_fractal_1996} identifies the level of the QS as the minimum of the average densities of disc regions with density values within $\pm k\sigma$, where $k$ takes values between 0.1 and 3.0.

\subsection{CLV extraction}
We compute the radially symmetric background (SB hereafter), which in turn gives the 1D QS CLV, by applying a 5th degree polynomial fit to the density values $d=f(\mu)$ of the deduced NSB. All image pixels are considered and the fit is weighted with the local $\sigma$ defined within 100 concentric and equal area annuli.
This is the most important step for the calibration process, because the way the 1D QS CLV is defined, it determines the CC.

Our calculation of NSB includes possible vignetting affecting the original SHG, however this leads to miscalculation of the 1D QS CLV.
Under-exposure depends on the intensity level, while vignetting is a purely geometric effect. Unfortunately, the geometry is such that their effects are similar, and vignetting is known to darken solar observations towards the limb, resembling the effects of under-exposure. The effects of vignetting are expected to be restricted mostly at the edges of the image, in  which case it might not affect the solar disc at all or it is almost indistinguishable from under-exposure.
Notwithstanding the similarities of their effects on the solar observation, some differences in the signatures of vignetting and under-exposure on the QS CLV and CC exist.  
In fact, vignetting can potentially affect larger parts of the solar observation than under-exposure, by modifying the CLV even near the image centre in contrast to under-exposure that mainly affects observations near the solar limb (see Fig. \ref{fig:synth5_characteristiccurves}). 
This is valid as long as the under-exposure is not extreme. However, in extreme cases the image quality is anyway very poor and such images cannot be processed meaningfully.

We attempt to estimate the vignetting on the analysed image and account for it in the calculation of the 1D QS CLV, by comparing the SB with a rescaled version of the logarithm of the CLV retrieved from the CCD-based Rome/PSPT observations. 
If there is no vignetting, the SB from the polynomial fit is kept. 
Otherwise, the SB is computed by rescaling the CLV from the modern observations. Its lower values are adjusted so as to find the minimum density for which the difference between SB and the CLV derived with the fit continuously increases towards the limb.

%%%%%%%%%%%%%%%%%%%%

\subsection{Photometric calibration}
\label{sec:photometriccalibration}
In our approach, we perform the photometric calibration by seeking for information stored on the solar disc that can be used in a similar manner as the often missing calibration wedges.

\begin{figure}[h!]
	\centering
	\includegraphics[width=1\linewidth]{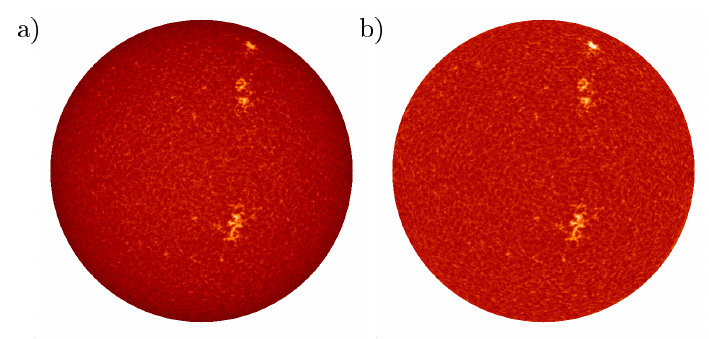}
	\caption{(a) Photometrically calibrated, and (b) contrast image corrected for the CLV for the Ko observation shown in Fig. \ref{fig:nsb_steps}. The colour scale of the calibrated image covers the full range of intensities on the disc, while the contrast image is shown within the intensity range [-1,1].}
	\label{fig:calib_steps}
\end{figure}

For each analysed image, we deduce the CC  by relating the density values obtained from its SB at a given $\mu$ position  to the logarithm of intensity values derived from modern Rome/PSPT observations at the same disc location. The amount of equal area annuli we use to acquire this relation and apply a linear fit to it is equal to the nearest integer of $2R$. This is  consistent with the assumption
that the QS density values in good-quality observations lie on the linear part of the CC.
From the fit we only exclude the last value that corresponds to the regions very close to the limb, due to the higher uncertainties that characterise these regions. 
We linearly extrapolate the computed relation to the whole range of density values on the image and make use of the fit parameters to photometrically calibrate the original density image. The result of this procedure is illustrated in Fig. \ref{fig:calib_steps}\textbf{a)}. We also photometrically calibrate the estimated NSB. Removal of the QS CLV from the calibrated image and normalization to it then provides the contrast image of the analysed SHG (Fig. \ref{fig:calib_steps}\textbf{b)}).

\begin{figure}[h!]	\includegraphics[width=1.0\linewidth]{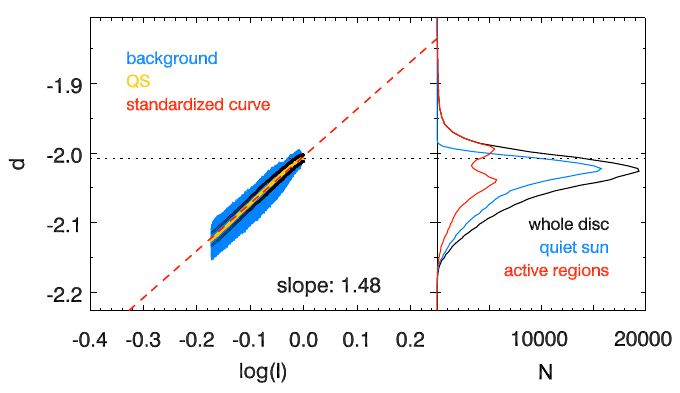}
	\caption{\textit{Left}: CC derived from the processing of the Ko observation shown in Fig. \ref{fig:nsb_steps} (red), measured CC for the QS (orange) with 1$\sigma$ uncertainty (black) and the whole background (blue). The slope of the derived CC is also shown. \textit{Right}: Distribution of density values for the QS (blue), AR (red) and whole disc (black). The horizontal dashed line in both panels denotes the highest value of the QS CLV.}
	\label{fig:03_00_calibrationcurve}	
\end{figure}

Figure \ref{fig:03_00_calibrationcurve} shows the CC derived from the processing of the Ko observation displayed in Fig. \ref{fig:nsb_steps}\textbf{a)}.

\section{Performance of the method}
\label{sec:results}

\subsection{Synthetic data}
\label{sec:testonsyntheticdata}
We tested the method proposed in Sect. \ref{sec:method} on a large number of synthetic images that were obtained from contrast Rome/PSPT images by imposing on them known radially symmetric CLV and a variety of inhomogeneities identified in historical observations (and converted to intensity). We then converted the degraded images from intensities to densities by applying various CC, thus emulating historical observations.
We used the 2000 Rome/PSPT images described in Sect. \ref{sec:data}, which were processed as described in Sect. \ref{sec:backgroundcomputation} to derive contrast images.
These contrast images are defined as $C^I_i=(I_i-I^{QS}_i)/I^{QS}_i$, where $I_i$ is the intensity, and $I_i^{QS}$ the intensity of the QS of the $i$th pixel.
The imposed CLV are in the form of a 5th degree polynomial function of $\ln{\mu}$, as presented by \cite{pierce_solar_1977}, with parameters determined from the Rome/PSPT observations. 
The range of values of the imposed intensity CLV is [0.6, 1.0], while the contrast range of the facular pattern is [0.1, $\sim$0.6], and of the network pattern is [0.026, 0.1].    
These ranges match the ones found on images of the Ko, Ar and Mi archives.\\

We produced 8 subsets of synthetic images with the following features:\\
\noindent {\it - Subset 1:} Ideal density images with symmetric backgrounds, no linear artefacts or bands, to estimate the precision of the proposed method and the sensitivity of results to the level of solar activity. \\
\noindent {\it - Subset 2:} Images with imposed CC that are linear functions with varying slopes, to test the sensitivity of the method to different slopes of CC within the range 0.1 -- 4.0. \\
\noindent {\it - Subset 3:} Images with imposed CC that are  non-linear functions  with various levels of over- and under-exposure (shown in Fig. \ref{fig:psptnonlinear}), to estimate errors in the retrieved calibration  
due to exposure problems. \\
\noindent {\it - Subset 4:} Images with different sizes, to test the accuracy of the method on the various datasets. The disc diameter was varied between 200 and 1100 pixels.\\ 
\noindent  {\it - Subset 5:} A vignetting function with different strength levels was added to images, to test the ability of the method to identify possible vignetting effect and to distinguish it from other artefacts. The vignetting used here is axi-symmetric.\\
\noindent {\it - Subset 6:} Various large scale density patterns of non-solar origin were imposed  (shown in Fig. \ref{fig:synth5_background}), to evaluate the efficiency of the method to distinguish between the solar and non-solar patterns and the accuracy of accounting for the latter.\\
\noindent {\it - Subset 7:} Images with a different CLV  (shown in Fig. \ref{fig:difclv}) than the one used during the standardization of the CC, to test the errors of calibrating data obtained with different bandwidths.\\
\noindent {\it -- Subset 8:} A combination of all the artefacts mentioned above was added randomly on each image, to produce an inhomogeneous dataset resembling the historical ones. The CC is defined as a 3rd degree polynomial function with randomly selected parameters within the following ranges: [-4.6, -1.3] for the constant term, [0.5, 4.0] for the linear term, [-0.03, 0.00] for the quadratic term, and [-0.03, 0.00] for the cubic term. The range of intensities within the undegraded images varies and the logarithm of it lies between -1.2 and 0.5 (similar to those shown with the light grey area in Fig. \ref{fig:psptnonlinear}).\\

\begin{figure*}[t!]
	\centering
	\includegraphics[width=1\linewidth]{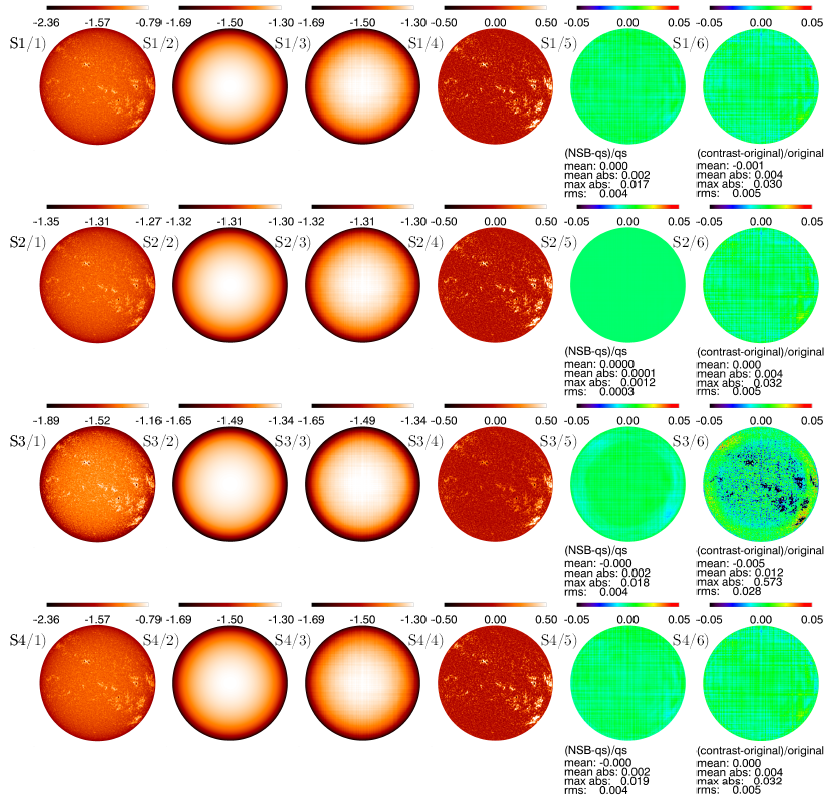}
	\caption{Examples of the calibration procedure on synthetic images from subsets 1 -- 4 (from top to bottom) produced by Rome/PSPT image taken on 21/08/2000. From left to right: (\textbf{1)}) density images, (\textbf{2)}) imposed backgrounds, (\textbf{3)}) calculated backgrounds, (\textbf{4)}) calibrated contrast images, (\textbf{5)}) pixel-by-pixel relative errors in NSB, and (\textbf{6)}) relative errors in calibrated contrast images. Here we show the following cases: subset 2 -- CC with slope of 0.1; subset 3 -- combination of the strongest under- and over- exposure (case 10 underexposure and 10 overexposure in Fig. \ref{fig:psptnonlinear}); subset 4 -- disc diameter of 1100 pixels. Also given (below the images in columns 5 and 6) are the values of the RMS, mean, mean absolute and maximum relative differences within the disc up to $0.98R$.} 
	\label{fig:synthexamples1}
\end{figure*}

\begin{figure*}[t!]
	\centering
	\includegraphics[width=1\linewidth]{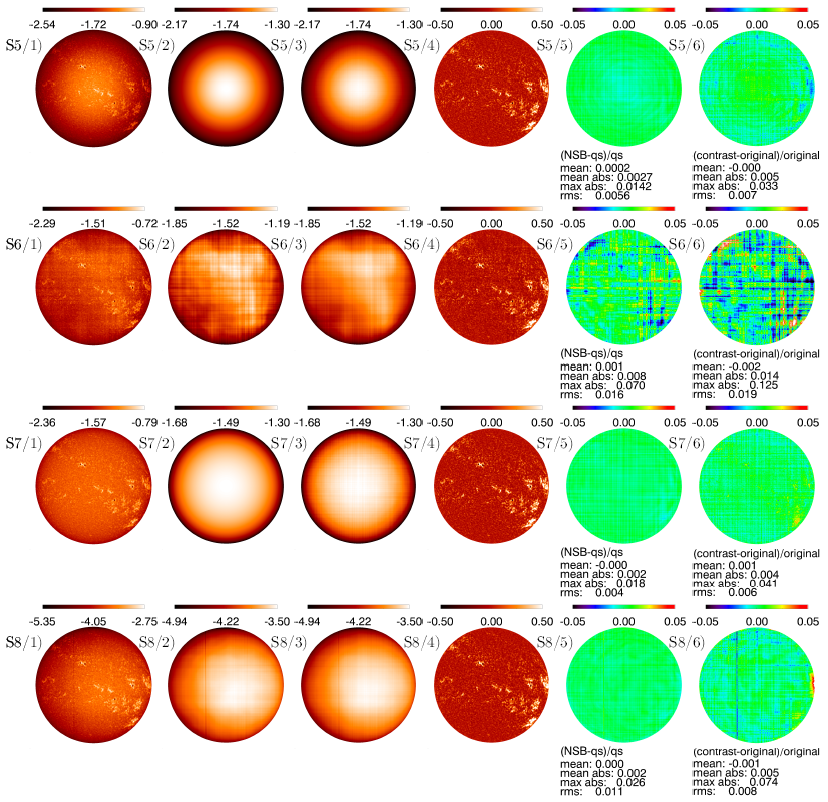}
	\caption{Same as Fig. \ref{fig:synthexamples1} but for subsets 5 -- 8. Here we show the following cases: subset 5 -- strongest vignetting (red curve in Fig. \ref{fig:synth5_characteristiccurves}); subset 6 -- level 5 of inhomogeneity No. 2 (shown in Fig. \ref{fig:synth5_background}); subset 7 -- CLV No. 4 (shown in Fig. \ref{fig:difclv}); subset 8 -- vignetting No. 2, level 5 of inhomogeneity No. 4, CC $d=-3.6+3.6\log{I}-0.01(\log{I})^2-0.004(\log{I})^3$, and CLV used $d=1.0+0.33\log(\mu)+0.06(\log(\mu))^2-0.04(\log(\mu))^3-0.05(\log(\mu))^4-0.01(\log(\mu))^5$ that lies roughly between cases 6 and 7 in Fig. \ref{fig:difclv}.} 
	\label{fig:synthexamples2}
\end{figure*}

Figures \ref{fig:synthexamples1} and \ref{fig:synthexamples2} show some examples of the synthetic images derived from the same contrast Rome/PSPT observation (taken on 21/08/2000), illustrating the variety of problems we aim at addressing with the application of the proposed method on these data. 
In the same figures we show the results obtained for these synthetic data and the pixel-by-pixel errors in the NSB calculation and the calibrated contrast images. The contrast images used to derive the errors are offset so that the mean QS contrast value is 1.

\subsubsection{Overall performance of our method on synthetic data}
\label{sec:overallsynthetic}
Table \ref{tab:syntheticdatasubsets} summarises key aspects of the various synthetic datasets, and quantitative results obtained by testing the proposed method on them. 
The results are briefly presented in the following and are described in greater detail in Appendix \ref{sec:results_synth_ap}.
Table \ref{tab:syntheticdatasubsets} summarizes the results derived from the analysis of all data within the various subsets. An exception is made for subsets 3 and 6, for which the presented values correspond to results restricted to images representative of historical data of reasonable quality, i.e. excluding synthetic images made to suffer from extreme exposure problems affecting the QS regions or images with superposed large-scale patterns whose amplitudes were larger than 0.6 of the CLV.

The results obtained for subsets 1--6 show that our method recovers the QS density CLV with average relative error $<3$\%  in the absence of strong non-solar patterns affecting the image, and error $<6.5$\% if the extreme cases of inhomogeneities encompassed in subset 6 are also included. 
The results derived from subset 7 prove that the above accuracy is maintained as long as the CLV differs by roughly $<10$\% from the one we imposed on the data.

\begin{table*}
	\caption{Summary of the characteristics of synthetic data and fidelity of retrieved images. }
	\label{tab:syntheticdatasubsets}
	\centering
	\begin{tabular}{lcccccccccc}
		\hline\hline
		Subset & N & Type of CC & N CC & Inhomogeneities & \multicolumn{2}{c}{NSB Max [\%]}	& \multicolumn{2}{c}{Contrast Max [\%]} & \multicolumn{2}{c}{Contrast RMS [\%]} \\
		(1)&(2)&(3)&(4)&(5)&\multicolumn{2}{c}{(6)} &\multicolumn{2}{c}{(7)} &\multicolumn{2}{c}{(8)} \\
		&  &  & & & mean & range &	mean & range & mean & range \\
		\hline
		1 & 500 & linear 	 & 1  & no        			& 1.3 &  0.8--2.3 & 2.4 &  1.5--4.0 &  0.5 &  0.3--1.0 \\
		2 & 200 & linear 	 & 20 & no         			& 1.3 &  0.1--3.1 & 3.1 &  1.7--5.2 &  0.6 &  0.3--0.8 \\
		3 & 1000& 3rd degree & 100& no     				& 1.3 &  0.8--1.8 & 6.4 &  1.5--25.1 &  0.8 &  0.3--2.4 \\
		4 & 100 & linear 	 & 1  & different disc sizes & 1.8 &  1.0--3.2 & 2.7 &  1.8--5.4 &  0.5 &  0.3--0.9 \\
		5 & 100 & linear 	 & 1  & vignetting   		& 1.4 &  0.9--1.8 & 2.6 &  1.5--4.0 &  0.6 &  0.3--0.8 \\
		6 & 2000& linear 	 & 1  & NSB      			& 2.9 & 1.1--11.3 & 5.1 & 1.9--32.5 &  0.9 & 0.4--5.9 \\
		7 & 100 & linear 	 & 1  & different CLV 		& 1.4 &  0.9--1.9 & 9.8 &  1.6--31.7 &  2.1 &  0.3--6.5 \\
		8 & 2000& 3rd degree &2000& all 				& 1.8 &  0.2--18.3 & 6.5 &  1.5--29.0 &  1.2 &  0.3--6.4 \\
		\hline                                            
	\end{tabular}
\tablefoot{(1) ID number of the subset, (2) number of images within the subset, (3) type and (4) number of different CC, (5) type of inhomogeneities, (6) relative maximum error in the retrieved NSB, (7) maximum relative and (8) RMS errors of the calibrated contrast images. For (6)--(8) we provide the average and the range of values in percent.}
\end{table*}

\label{sec:results_synth8_randomproblems}
The results obtained from the subset 8 demonstrate that the proposed method retrieves the NSB affecting the observation with a maximum relative error $<2\%$ averaged over all the analysed images.
We found that, similarly to artefacts on the historical images, the presence of gradients on the synthetic images introduces errors in the calculation of SB, which affects the range of values in the calibrated data. 

\begin{figure}
	\centering
	\includegraphics[width=1\linewidth]{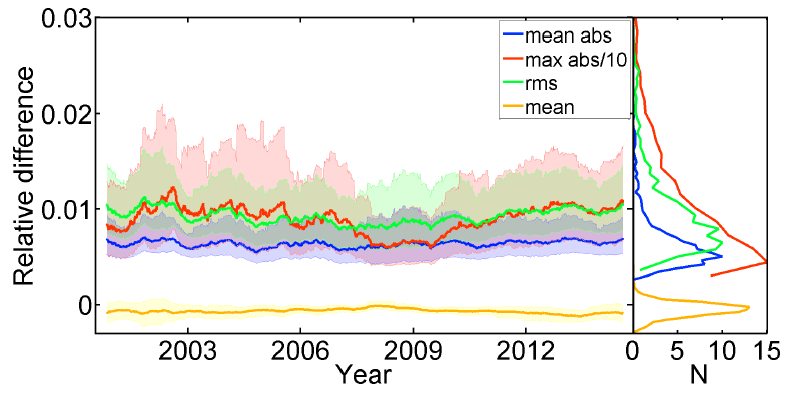}
	\caption{{\it Left}: Relative difference between the calibrated and CLV-corrected image with our method and the original image (within $0.98R$) for all the synthetic data of subset 8. RMS difference (green), mean absolute difference (blue), mean difference (orange) and maximum difference (red). Each of these values refers to a single image at a time (e.g. the difference averaged over all pixels, or the maximum value found in one pixel of the image). These differences are plotted vs. the date on which the original Rome/PSPT images (that were randomly distorted) were recorded. Note that the maximum difference values have been divided by 10 to plot them on the same scale as the other quantities. The solid lines are 100 point averages (i.e. averages over the values obtained for 100 images) and the shaded surfaces denote the asymmetric 1$\sigma$ interval.  {\it Right}: Distribution of the relative difference values.}
	\label{fig:synth8_flat_max98}
\end{figure}

Figure  \ref{fig:synth8_flat_max98} illustrates the accuracy of the proposed method on the most challenging subset 8. Shown are the maximum and average values of the relative difference as well as the RMS difference between the calibrated and processed contrast image and the original undegraded synthetic contrast image. Throughout this analysis the relative differences are given in absolute values (unsigned), but we also provide the signed average difference.
These values quantify the final errors of our image processing, which is comprised of the photometric calibration and the removal of the CLV and other image patterns from the analysed image and finally provides the corrected contrast image. 
We found that the maximum differences are on average $<6.5$\%, while the average differences are $<1$\%.  
There is, however, a tail consisting of images with higher maximum or average differences. For images representative of low activity periods, maximum errors are approximately 2\% lower than those found for images at high activity periods, which illustrates the need for a very careful removal of active areas prior to carrying out any image processing. We discern no significant variation of the mean and RMS differences over the solar cycle.

\begin{figure}
	\centering
	\includegraphics[width=1.0\linewidth]{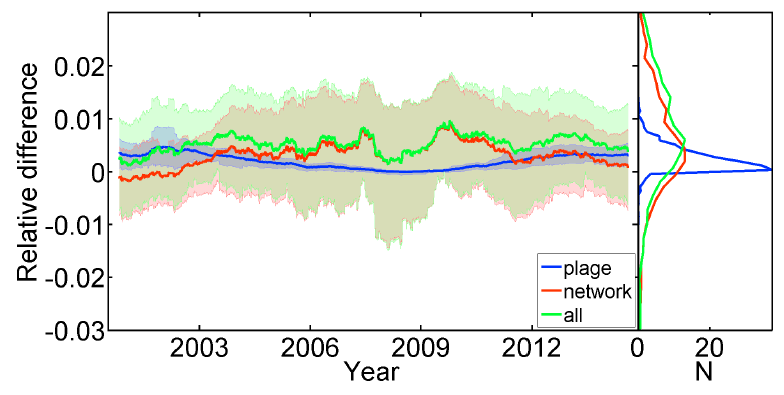}
	\caption{{\it Left}: Relative difference between the disc fractions of AR obtained on the images processed with our method and on the original undegraded images of subset 8. Disc fractions are shown separately for plage (blue), and network (red), as well as for their sum (green). The features were identified with constant thresholds. The solid lines are 100 point averages and the shaded areas show the asymmetric 1$\sigma$ intervals.  {\it Right}: Distribution of the relative difference values.}
	\label{fig:synth8_residualdiscfractions}
\end{figure}

We also tested the accuracy of our proposed method for studies of variations in the fractional coverage of the solar disc by AR and network.
Figure \ref{fig:synth8_residualdiscfractions} shows the relative difference between the disc fraction of various solar features identified on the processed and unprocessed images of subset 8. 
The disc features were identified by applying a set of constant thresholds on the original (i.e. prior to degradation) and final (i.e. after recovery) contrast images. The thresholds used here were defined on the Rome/PSPT data to describe the plage and network regions. The thresholds in the contrast values used are $th_p=0.21$ and $th_n=0.03$ for the plage and network, respectively.
Figure \ref{fig:synth8_residualdiscfractions} demonstrates that the estimated disc fractions for the processed data typically lie within 2\% of the values derived from the original data.

\subsubsection{Performance on individual steps of the processing}
\label{sec:NSB calculation}
Our method includes original ideas for processing SHG (e.g.  rotating slices), but it also partly uses ideas from the previously published methods described in Sect. \ref{sec:intro}. By testing all these methods on synthetic data we identified their drawbacks, which helped us to optimize the steps of the method proposed here.

For example, in our calculation of the background, we do not rotate the image by $45^\circ$ as proposed by \citet{worden_evolution_1998}. 
We found that the rotation does not improve the accuracy of the image processing further, if the outcome of other processing steps has been optimized. 
In contrast, our iterative fitting improves the accuracy of the AR identification and, in turn, of the QS estimation as compared to both non-iterative computations with a 5th degree polynomial function as suggested by \citet{worden_evolution_1998} and iterative computations with higher degree functions. 
However, we also noticed that on average more than three computations of the fit per iteration step merely results in an increase of the noise of the final NSB map derived from the processing without improving the accuracy of the result.

In \textit{Step 2.2}, to identify and exclude AR we apply a thresholding scheme with asymmetric limits, by using the values $^{+0.5}_{-1}\sigma$ instead of the more widely employed  $\pm 2\sigma$. 
The asymmetric range allows us to account for both, the potentially inaccurate identification of AR near the solar limb at earlier processing steps and the small disc fraction of dark features in Ca~II~K observations. 
For example, using the symmetric limits of $\pm 1 \sigma$ on the synthetic image in Fig.  \ref{fig:synthexamples1} \textbf{S1/1)}  results in increased errors in the NSB of 3.5\% and overestimation of AR intensity values (the errors with our method and the asymmetric limits are 1.8\% and are shown in Fig. \ref{fig:synthexamples1} \textbf{S1/5)}. 
The employed upper limit of $+0.5$\% was found to be a good compromise. Larger values led to the inclusion of significant portions of AR in the QS background, while lower values for the lower limit risk to wrongly exclude large regions of the solar observations near the limb from the 2D QS background calculation.

\begin{figure}[h!]
	\centering
	\includegraphics[width=1\linewidth]{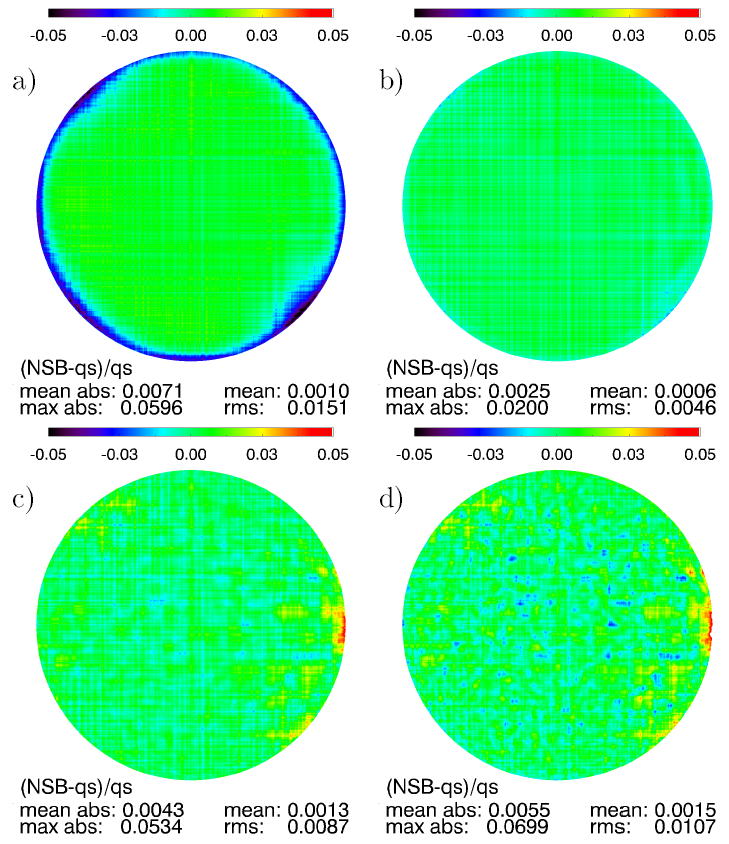}
	\caption{Relative error in NSB calculation for one image of subset 1 derived from Rome/PSPT observation taken on 21/08/2000 shown in Fig.  \ref{fig:synthexamples1} \textbf{S1/1)}. The NSB was derived with our method and running window median filter width of (\textbf{a)}) $R/2$,  (\textbf{b)}) $R/4$, (\textbf{c)}) $R/13$, and (\textbf{d)}) $R/20$ pixels. Also shown are the values of the RMS, mean, mean absolute, and maximum relative differences by comparing image regions within 0.98$R$. The colour bars apply to the images below them. }
	\label{fig:examplesNSBwindowwidths}
\end{figure}

The window width for the median filter used in \textit{Step 2.4} was chosen to scale with $R$, to achieve consistent results from different data with varying disc size. 
The adopted width is larger than the typical scale of the network on the analysed observations, in order to avoid effects of small-scale density patterns of solar origin on image processing results, but is small enough to account for rapid changes of the background near the solar limb. 
We found that window widths in the range $R/6 - R/8$ perform best on all available data and we adopted the more conservative value of $R/6$.
This finding is in contrast to that of \citet{chatterjee_butterfly_2016} who used a window width of $\sim R/13$.
Figure \ref{fig:examplesNSBwindowwidths} shows one example of testing different window widths on an image from subset 1 (shown in Fig.  \ref{fig:synthexamples1} \textbf{S1/1)}).
We show widths of $R/2$, $R/4$, $R/13$, and $R/20$ pixels.
Large window widths fail very close to the limb, while smaller widths progressively fail on AR and network. Window width of $R/4$ gives comparable results with those of our adopted value, but slightly underestimates the values very close to the limb.

The accuracy of our processing also decreases if we do not identify the density variations of adjacent image lines\footnote[2]{Such linear artefacts may have been introduced during the observation due to problems of the spectroheliograph employed, e.g. irregular drive.} and simply derive a smooth background map. 
Furthermore, if these variations are not properly accounted for in the analysed image, its subsequent analysis aimed at the estimation of e.g. the photometric properties of AR returns inaccurate results.

Tests on subset 3 that includes non-linear CC, showed that our method is very accurate in recovering the shape of the CC even on observations with strong exposure problems, with  relative errors in the computed CC 
being usually $<0.5$\% under typical conditions as well as for the other synthetic subsets. Even in the extreme cases of over- or under-exposure the relative error in the computed CC lies below 1.7\%. 

\begin{figure}[h!]
	\centering
	\includegraphics[width=1\linewidth]{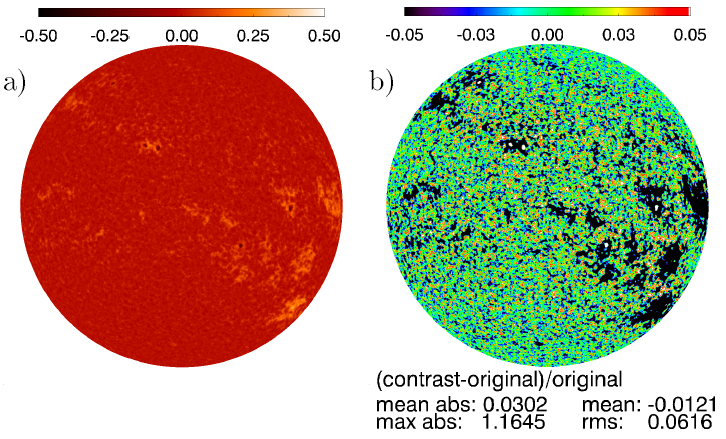}
	\caption{Examples of applying our method on a synthetic image of subset 5 derived from Rome/PSPT observation taken on 21/08/2000: (a) contrast image derived after processing of the synthetic image without compensation for the vignetting; (b) relative error of the calibrated contrast images. Also shown are the values of the RMS, mean, mean absolute, and maximum relative differences by comparing image regions within 0.98$R$. The colour bars apply to the images below them.}
	\label{fig:examplesvignetting}
\end{figure}

The analysis of the images of the subset 2 suggests that the CC slope only mildly affects the results. Our method also accurately disentangles the vignetting contribution on the CC (see Fig. \ref{fig:synth5_characteristiccurves}). Figure \ref{fig:examplesvignetting} shows an example of processing an image from subset 5 without attempting to recover the vignetting (see Fig. \ref{fig:synthexamples2} for the results with the vignetting recovery). The vignetting contribution increases the slope of the CC and reduces the contrast values of the plage regions, with the errors over these regions being 35 times greater than if we account for the vignetting with our method.

\begin{figure*}[t!]
	\centering
	\includegraphics[width=1\linewidth]{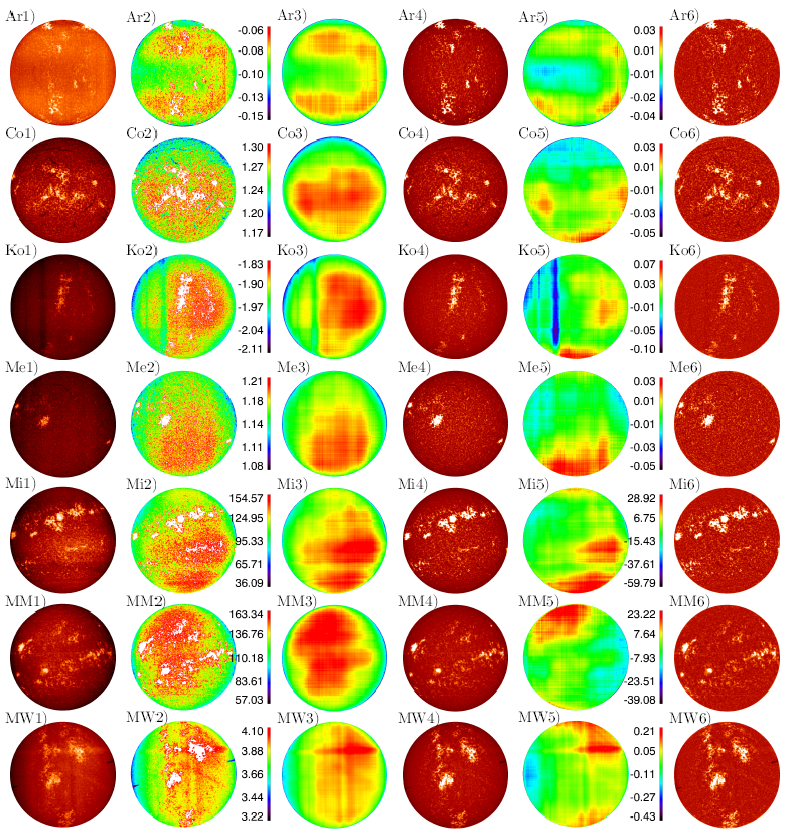}
	\caption{Examples of the calibration procedure of historical images from the Ar, Co, Ko, Me, Mi, MM, MW archives (from top to bottom) taken at high activity periods. From left to right: (1) density images, (2) density images saturated such as to clearly show the backgrounds, (3) calculated backgrounds (CLV and inhomogeneities), (4) calibrated images, (5) identified inhomogeneities, and (6) images corrected for QS CLV. All the unprocessed density images are shown with the whole range of densities within the disc, the calibrated images are shown within the intensity range [0.0, 2.0], while the contrast images (i.e. images compensated for the CLV) are plotted within [-0.5, 0.5]. A colour bar gives the density/intensity range for the rest of the images. The colour bar between images (2) and (3) applies to both images.} 
	\label{fig:shgexamples1}
\end{figure*}

\begin{figure*}[t!]
	\centering
	\includegraphics[width=1\linewidth]{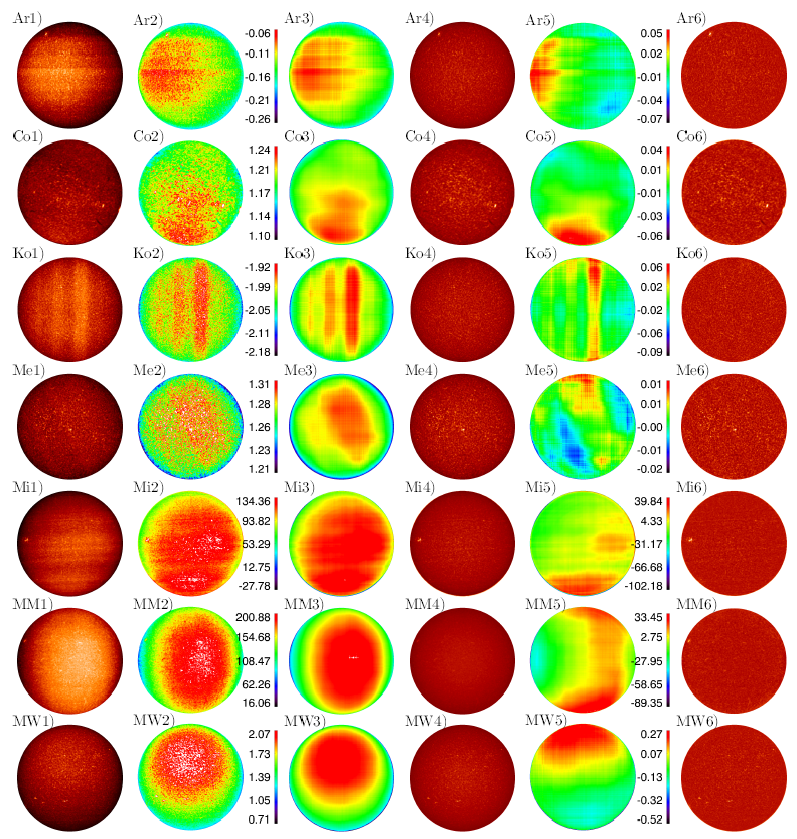}
	\caption{As Fig. \ref{fig:shgexamples1}, but for images taken at low activity periods.} 
	\label{fig:shgexamples2}
\end{figure*}

\subsection{Examples of calibrated SHG}
\label{sec:results_shg}
We have applied the proposed method to many SHG randomly selected from the seven available historical (photographic) archives. 
Figures \ref{fig:shgexamples1} and \ref{fig:shgexamples2} show examples of the results obtained from observations taken at periods of high and low solar activity, respectively. 
From left to right, each panel shows  the original observation (density image), the density image saturated to show the background,
the background (CLV plus inhomogeneities) deduced from the proposed processing,  the calibrated image,  the identified inhomogeneities of the analysed image, and the final contrast image after the QS CLV removal. From top to bottom each such set of images is shown for SHG extracted from the Ar, Co, Ko, Me, Mi, MM, and MW archives, respectively. 
All the calibrated images are shown within the intensity range [0.0, 2.0], while the contrast images (i.e. images compensated for the CLV) are plotted within [-0.5, 0.5]. For the rest there is a colour bar denoting the range of values.
Further examples can be found in \cite{chatzistergos_exploiting_2016}.

The deduced backgrounds describe the different patterns in the analysed images quite accurately.
Figures \ref{fig:shgexamples1} and \ref{fig:shgexamples2} clearly show that the method works consistently with data extracted from various photographic archives, taken at different activity levels.
All calibrated images lie within the same range of values and show a similar CLV pattern.
The same is true for the contrast images that return plage regions within the same intensity ranges and no obvious residual large scale artefacts. 
The method is able to account even for strong inhomogeneities and rather peculiar patterns (e.g. Fig. \ref{fig:shgexamples1} Ar, or Fig. \ref{fig:shgexamples2} Ko) without affecting the plage regions.
The inhomogeneities identified here show a CLV that is usually off-centred, in many cases having its highest value towards the limb.
Furthermore, images show many dark/bright bands that could occur due to something occluding the Sun for a short period, or not constant exposure over the different rasters.

The tests on historical observations with good quality, resulted in 1D QS CLV very similar to the one from the Rome/PSPT data. This strengthens our argument that the CC of the good quality historical data can be described by a linear relation.

\section{Comparison with other methods}
\label{sec:comparisonwithothermethods}
\subsection{Background calculation methods}
\label{sec:discussion}

\label{sec:discussion_01_nsb}
We found that the methods presented in the literature that apply radially symmetric computations suffer from their inability to account for the asymmetric patterns affecting the images, while application of median filtering suffers under the presence of AR and its inability to account for density variations along adjacent image lines. 

By applying the methods of \cite{caccin_variations_1998} and \cite{worden_plage_1998} on the degraded synthetic SHG we found that these techniques provide inaccurate QS CLV values towards the limb and do not account for small-scale image artefacts, e.g. the density variations in adjacent lines. 
The method by \citet{tlatov_new_2009} is also unable to account for these linear artefacts.

\begin{figure}[h!]
	\centering
	\includegraphics[width=1\linewidth]{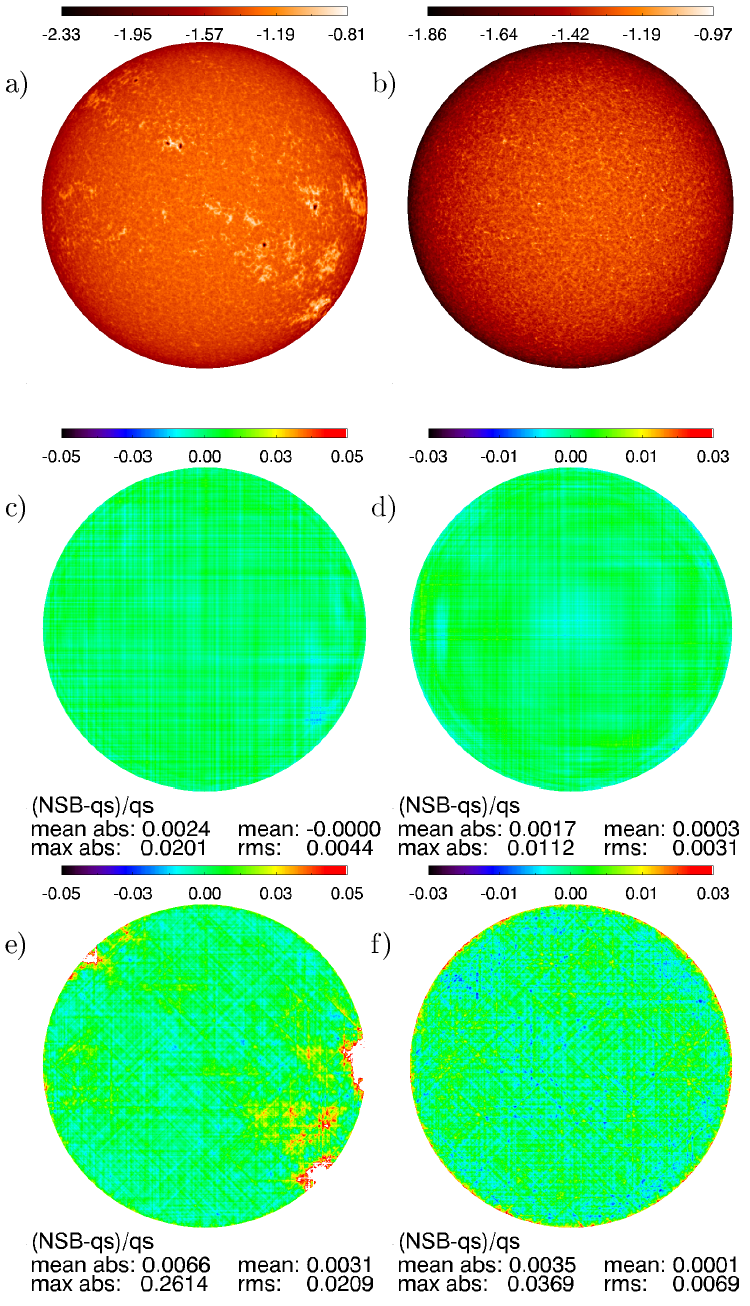}
	\caption{Relative error in NSB calculation with our method (c--d) and the method by \citet[][e--f]{worden_evolution_1998} for two images of subset 6 with the lowest level of inhomogeneities (density images shown on a--b). The facular pattern was derived from the Rome/PSPT observations taken on 21/08/2000 (left) and 09/08/2008 (right). Also given (below the images) are the values of the RMS, mean, mean absolute and maximum relative differences within the disc up to $0.98R$.  The colour bars apply to the images below them. }
	\label{fig:comparison_worden}
\end{figure}

Figure \ref{fig:comparison_worden} shows the pixel by pixel relative differences between the NSB and the imposed background derived with our method and that by \citet{worden_evolution_1998}\footnote[3]{When we applied the method by \cite{worden_evolution_1998} we did not perform the last step of the low-pass filtering, because the information of the window-width or the way they applied it on regions very close to the limb is not described. Applying this step could potentially reduce the errors of isolated pixels, but it would not make a difference in the misidentification of the AR.}, on two synthetic images from subset 6. 
Part of the AR remained undetected by the latter method and so enters the computation of the background. 
Thus the method by \cite{worden_evolution_1998} overestimates the actual background in some plage areas and introduces  processing errors in recovering the NSB. 
The maximum relative errors in NSB are lower for observations taken at low solar activity (3.7\%), than at periods of high solar activity (26\%), but on average the method by \cite{worden_evolution_1998} introduces one order of magnitude higher errors over the disc than obtained from our proposed method.

We applied the method by \citet{priyal_long_2013} on all data of subset 6 
and we found that the method by \citet{priyal_long_2013} works reasonably well on images with weak anisotropies, but consistently fails to account for the large inhomogeneities affecting the data, by introducing up to 16 (25) times larger maximum (RMS) errors than those from our method. 
Besides, the method by \citet{priyal_long_2013} does not allow recovering any image patterns that occur in a direction different from the one considered for the fit.

\subsection{Calibration methods}
\begin{figure}
	\centering
	\includegraphics[width=1.0\linewidth]{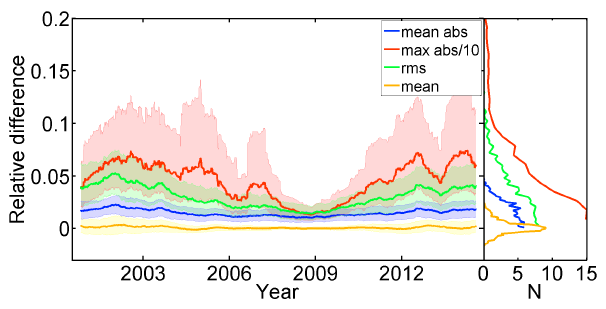}
	\caption{{\it Left}: Relative error for the contrast images derived from the calibration with the average CC as suggested by \cite{priyal_long_2013} of all the synthetic images of subset 8. The maximum differences are divided by 10. Note that the red line (maximum value of the unsigned relative difference) has been divided by 10 to allow it to be plotted together with the other curves. The solid lines are 100 point averages and the shaded surfaces denote the asymmetric 1$\sigma$ interval.  {\it Right}: Distribution of the relative difference values.}
	\label{fig:synth9_flat}
\end{figure}

We tested the accuracy of the method by \cite{priyal_long_2013} by applying an average CC to calibrate a whole dataset with our synthetic data.
We derived the average CC from the curves we imposed to all the data of subset 8, and studied the differences between contrast images obtained from the original data  and the ones resulting from the calibration with the average CC. 
We used the imposed background of each image to compensate for the limb darkening, in order to avoid any other uncertainties of our procedure. 
This error estimate can be considered only as a lower limit, since the CC used to derive the average were the imposed ones, therefore without taking into consideration any errors in the calculation of the individual CC.
Figure \ref{fig:synth9_flat} shows that the errors introduced by the calibration with the average CC as proposed by the method by \citet{priyal_long_2013} are on average $\sim$50\%. These errors reach values as high as 300\% for a few cases. 
We stress, however, that we cannot rule out that in actual historical data sets the CC display a smaller variation than in our subset 8. Therefore, the above test mainly shows the greater versatility of the present technique for handling a range of CC values. 

\begin{figure}[h!]
	\centering
	\includegraphics[width=1\linewidth]{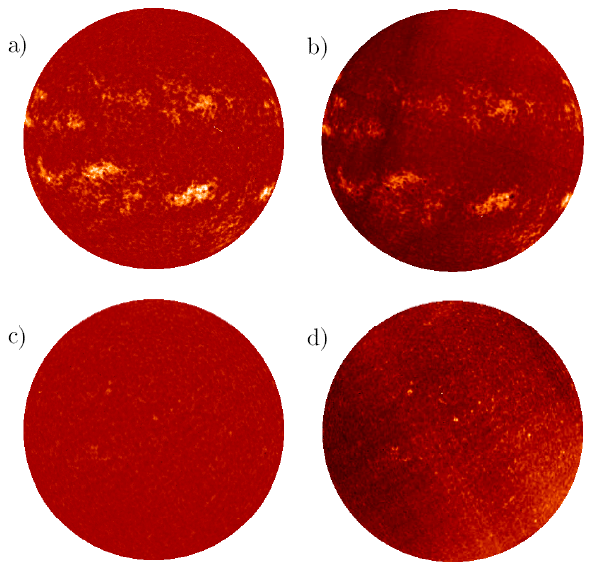}
	\caption{Examples of calibrated and CLV compensated images from the Ko archive derived with our method (left) and with the method of \citet[][right]{priyal_long_2013};  the latter data were taken from the Kodaikanal website. The results from \citet[][right]{priyal_long_2013} are given in JPG files and shown here unsaturated, while the results from our processing are saturated in the range [-0.5, 0.6]. The observations were taken on 05/08/1947 (top) and 01/01/1964 (bottom).}
	\label{fig:priyalcalibrated}
\end{figure}

Figure \ref{fig:priyalcalibrated} shows two examples of Ko images processed with our method (left panels) and by \cite{priyal_long_2013}\footnote[4]{Available at  \url{http://kso.iiap.res.in/data}}.
The input data employed for this comparison come from different digitizations of the same observation. 
This limits our analysis of the results to qualitative aspects only. 
Our images were saturated at the same level to illustrate all AR clearly, however the data by \cite{priyal_long_2013} were provided in JPG file format and hence we are unable to saturate the images to the same level. 
Still, Fig. \ref{fig:priyalcalibrated} clearly shows that images processed by \cite{priyal_long_2013} are affected by uncorrected inhomogeneities, to a significantly larger extent than images processed by our method.

\begin{figure}[h!]
	\centering
	\includegraphics[width=1\linewidth]{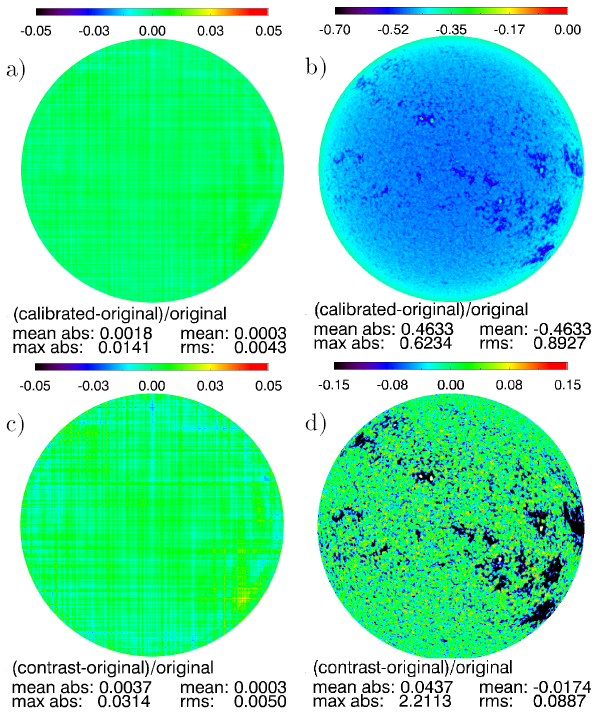}
	\caption{Relative error of the calibrated (top) and the contrast images (bottom) produced after linear calibration with our method (left) and the method of \citet[][right]{tlatov_new_2009} for an image of subset 1. The facular pattern was derived from a Rome/PSPT observation taken on 21/08/2000 (shown in Fig. \ref{fig:synthexamples1} S1). Also listed are the values of the RMS, mean, mean absolute and maximum relative differences by comparing image regions within $0.98R$. The colour bars apply to the images below them and are different for each image.}		
	\label{fig:comparison_tlatov}
\end{figure}

In the image processing suggested by \citet{tlatov_new_2009}, all image values are termed intensities and are further scaled linearly to the values of the standardized profile, without conversion to density values as expected by photographic theory, therefore this method merely applies a linear scaling to the image to let it match a desired range of intensity values. Also the standardised CLV that is taken from \citet{pierce_solar_1977} does not accurately represent the Ca~II~K data since it corresponds to 390.928 nm.
Figure \ref{fig:comparison_tlatov} shows the relative difference between results derived from the application of the method by  \citet{tlatov_new_2009}  and our method to one synthetic image from subset 1 produced from the contrast Rome/PSPT observation shown in Fig. \ref{fig:synthexamples1} S1. 
The image calibrated with the method of \citet{tlatov_new_2009} displays a significant offset {$\sim$0.7} and fainter plage and dark regions than in the original image \citep[contrast of $\sim$0.1 and $\sim$0.5 obtained by the method of][and ours, respectively]{tlatov_new_2009}.

The method by \cite{ermolli_comparison_2009} cannot be tested on the synthetic data, since it relies on information of the unexposed regions of the plate that we cannot replicate in a meaningful way in the synthetic data.

Finally, we compared the CC computed with our technique and from the calibration wedges stored on Ar data. 

Since our method sets the QS near the solar disc centre to be around 1, whereas the wedges describe the response of the whole plate and contain no information as to which range the QS corresponds, there is a scaling factor between the images calibrated with our method and with the wedges. 
This factor depends mostly on the digitization (i.e. the range of values of the QS in the digital files), but also on other factors (e.g. slightly different exposure time should change the location of the QS in the CC). 
Thus, a direct comparison of the CC derived from the two methods is not straightforward.

To account for the difference in values range, we derived the CC from the wedges by applying a polynomial fit \citep{ermolli_digitized_2009}, used this CC to calibrate the CLV calculated with our method and then rescaled it to match the range of values in the calibrated CLV derived with our method. 
Since information is lost with the rescaling, this approach does not allow any conclusion on the slope of the CC. 
Nonetheless, in this way we can test the assumption of using a linear curve to calibrate these data, provided the fitting of the wedge measurements is done accurately enough. 
However, this may not happen for all Ar data, because of insufficient or inaccurate information stored on the wedges.
The wedges of the Ar data usually consist of 3 scans for 7 known exposures, giving 7 points in intensities to fit the sigmoid CC. These values do not necessarily cover the whole range of values on the disc, or even if they do their number may be too low to describe a sigmoid function.

\begin{figure}
	\centering
	\includegraphics[width=1.0\linewidth]{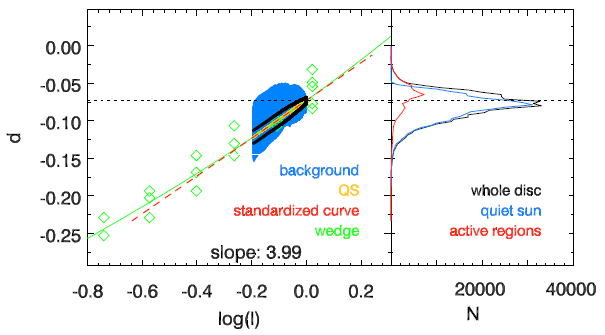}
	\caption{\textit{Left}: Standardized CC derived from our method (red, extrapolated to the range of values of the whole disc), measured CC for the QS (orange) with 1$\sigma$ uncertainty (black), the whole background (blue), calibration wedge measurements and fit (green rhombuses and line, respectively)  of  Ar observation taken on 20/07/1948. Shown also is the slope of the derived CC. \textit{Right}: Distribution of densities for the QS (blue), AR (red) and whole disc (black). The horizontal dashed line in both panels denotes the highest value of the QS CLV.}
	\label{fig:arcetriwedge}
\end{figure}

Figure \ref{fig:arcetriwedge} shows an example of a derived CC for one Ar observation including the rescaled wedge measurements and fit. 
This is one of the good cases where the CC derived with our method matches almost perfectly the one derived from the wedges, missing only a small part of underexposed regions. 
It is important to note that the scatter of the values derived from the wedges is almost the same as the scatter in the background of the image.
We achieve pixel by pixel relative differences between the image calibrated with our method and that with the wedge that are $<5$\%, which is consistent with the results presented with the synthetic data from subset 6 (the example tested here was found to have inhomogeneities with a range within 0.6 that of the CLV).

\section{Conclusions}
\label{sec:conclusions}
We have developed a new method to photometrically calibrate the historical full-disc Ca~II~K SHG and to correct them for various artefacts of solar and non-solar origin. 
The method is based on the standardization of the QS CLV intensity pattern to the one resulting from modern observations, under the assumption that it does not vary with time.
Modern observations suggest that this holds within the accuracy of the proposed method. 
We showed that the errors introduced by the above assumptions are relatively small and have minor impact on the CLV estimation unless the analysed observation is of very poor quality.
We assume that QS regions store all the information required to construct the CC for the range of brightnesses covered by the QS anywhere on the solar disc. 
This is not fulfilled for observations with strong over-exposure effects, as these introduce errors into the bright plage regions near the centre of the solar disc, that cannot be calibrated away by our method. Therefore very poor quality observations must be rejected. However, they constitute a very small fraction of the available historical data.
In addition, it can be that only part of the characteristic curve is accurately represented by the QS density values and the assumed linear relation.
This would affect mostly plage regions near disc centre or fainter regions towards the limb and can limit the accuracy of our processing.
However, we showed that in spite of these potential shortcomings, our method results in much lower errors than different approaches presented in the literature. 

To test the accuracy of the proposed method, we created a large number of synthetic images emulating various problems encountered in historical observations.
The maximum error of our method is $<6.5$\% averaged over all the degradations studied here, while the average error is $<1$\%. 
These errors were derived on synthetic data including extreme cases of imposed artefacts. The maximum errors reduce to $<2$\% if we exclude images with the most extreme artefacts.
Application of other methods for the processing of SHG presented in the literature, returns errors that are between 3 and 300 times larger than those derived from our method.

We also estimated the accuracy of processing modern Ca~II~K data by applying the proposed method to synthetic images unaffected by linear artefacts. 
The error estimates decreased by almost a factor of 2 with respect to those reported earlier, with maximum relative errors being on average $<0.6$\%.

Finally, we applied for test purposes the proposed method to a sample of images randomly extracted from seven historical SHG archives. 
We showed that the method allows us to process images from different archives consistently, without having to adjust or tailor the method to observations taken with different instruments at various observatories and at times of different levels of solar activity. 
The results of the application of the proposed method to various SHG archives will be presented in a forthcoming paper.

It is worth noting that our method to derive the CLV can be applied with minor adjustments to full-disc solar observations taken in other spectral ranges than Ca~II~K. Examples are archival white-light photographic images used for identifying and measuring sunspot properties \citep[e.g.][]{ravindra_digitized_2013,willis_greenwich_2013,hanaoka_long-term_2013}, or SHG in the H$\alpha$ line \citep[e.g.][]{mein_spectroheliograms_1990,potzi_scanning_2008,garcia_synoptic_2011,hanaoka_long-term_2013}.

\begin{acknowledgements}
We thank Dr. Tlatov for providing data processed with his method and for providing feedback on his method of calibrating SHG. 
The authors thank the Arcetri, Coimbra, Kodaikanal, McMath-Hulbert, Meudon, Mitaka, Mt Wilson, and the Rome Solar Groups.
T.C. thanks the INAF Osservatorio astronomico di Roma solar physics group for their hospitality during his lengthy visits.
T.C. acknowledges postgraduate fellowship of the International Max Planck Research School on Physical Processes in the Solar System and Beyond.
This work was supported by grants PRIN-INAF-2014 and PRIN/MIUR 2012P2HRCR "Il Sole attivo", COST Action ES1005 "TOSCA", FP7 SOLID, and by the BK21 plus program through the National Research Foundation (NRF) funded by the Ministry of Education of Korea.
\end{acknowledgements}

\bibliographystyle{aa}
\bibliography{_biblio1}   

\begin{thebibliography}{71}
\expandafter\ifx\csname natexlab\endcsname\relax\def\natexlab#1{#1}\fi

\bibitem[{Babcock \& Babcock(1955)}]{babcock_suns_1955}
Babcock, H.~W. \& Babcock, H.~D. 1955, The Astrophysical Journal, 121, 349

\bibitem[{Ball {et~al.}(2012)Ball, Unruh, Krivova, Solanki, Wenzler, Mortlock,
  \& Jaffe}]{ball_reconstruction_2012}
Ball, W.~T., Unruh, Y.~C., Krivova, N.~A., {et~al.} 2012, Astronomy and
  Astrophysics, 541, A27

\bibitem[{Bertello {et~al.}(2010)Bertello, Ulrich, \&
  Boyden}]{bertello_mount_2010}
Bertello, L., Ulrich, R.~K., \& Boyden, J.~E. 2010, Solar Physics, 264, 31

\bibitem[{Bühler {et~al.}(2013)Bühler, Lagg, \& Solanki}]{buhler_quiet_2013}
Bühler, D., Lagg, A., \& Solanki, S.~K. 2013, Astronomy and Astrophysics, 555,
  A33

\bibitem[{Brandt \& Steinegger(1998)}]{brandt_determination_1998}
Brandt, P.~N. \& Steinegger, M. 1998, Solar Physics, 177, 287

\bibitem[{Caccin {et~al.}(1998)Caccin, Ermolli, Fofi, \&
  Sambuco}]{caccin_variations_1998}
Caccin, B., Ermolli, I., Fofi, M., \& Sambuco, A.~M. 1998, Solar Physics, 177,
  295

\bibitem[{Chapman {et~al.}(2013)Chapman, Cookson, \&
  Preminger}]{chapman_modeling_2013}
Chapman, G.~A., Cookson, A.~M., \& Preminger, D.~G. 2013, Solar Physics, 283,
  295

\bibitem[{Chatterjee {et~al.}(2016)Chatterjee, Banerjee, \&
  Ravindra}]{chatterjee_butterfly_2016}
Chatterjee, S., Banerjee, D., \& Ravindra, B. 2016, The Astrophysical Journal,
  827, 87

\bibitem[{Chatzistergos {et~al.}(2016)Chatzistergos, Ermolli, Solanki, \&
  Krivova}]{chatzistergos_exploiting_2016}
Chatzistergos, T., Ermolli, I., Solanki, S.~K., \& Krivova, N.~A. 2016, in
  Astronomical {Society} of the {Pacific} {Conference} {Series}, Vol. 504,
  Coimbra {Solar} {Physics} {Meeting}: {Ground}-based {Solar} {Observations} in
  the {Space} {Instrumentation} {Era}, ed. I.~Dorotovic, C.~E. Fischer, \&
  M.~Temmer, San Francisco, 227--231

\bibitem[{Dainty \& Shaw(1974)}]{dainty_image_1974}
Dainty, J.~C. \& Shaw, R. 1974, Image science. {Principles}, analysis and
  evaluation of photographic-type imaging processes (London: Academic Press)

\bibitem[{Dasi-Espuig {et~al.}(2014)Dasi-Espuig, Jiang, Krivova, \&
  Solanki}]{dasi-espuig_modelling_2014}
Dasi-Espuig, M., Jiang, J., Krivova, N.~A., \& Solanki, S.~K. 2014, Astronomy
  and Astrophysics, 570, A23

\bibitem[{Dasi-Espuig {et~al.}(2016)Dasi-Espuig, Jiang, Krivova, Solanki,
  Unruh, \& Yeo}]{dasi-espuig_reconstruction_2016}
Dasi-Espuig, M., Jiang, J., Krivova, N.~A., {et~al.} 2016, Astronomy and
  Astrophysics, 590, A63

\bibitem[{Delaygue \& Bard(2011)}]{delaygue_antarctic_2011}
Delaygue, G. \& Bard, E. 2011, Climate Dynamics, 36, 2201

\bibitem[{Denker {et~al.}(1999)Denker, Johannesson, Marquette, Goode, Wang, \&
  Zirin}]{denker_synoptic_1999}
Denker, C., Johannesson, A., Marquette, W., {et~al.} 1999, Solar Physics, 184,
  87

\bibitem[{Ermolli {et~al.}(2007)Ermolli, Criscuoli, Centrone, Giorgi, \&
  Penza}]{ermolli_photometric_2007}
Ermolli, I., Criscuoli, S., Centrone, M., Giorgi, F., \& Penza, V. 2007,
  Astronomy and Astrophysics, 465, 305

\bibitem[{Ermolli {et~al.}(2011)Ermolli, Criscuoli, \&
  Giorgi}]{ermolli_recent_2011}
Ermolli, I., Criscuoli, S., \& Giorgi, F. 2011, Contributions of the
  Astronomical Observatory Skalnate Pleso, 41, 73

\bibitem[{Ermolli {et~al.}(1998)Ermolli, Fofi, Bernacchia, Berrilli, Caccin,
  Egidi, \& Florio}]{ermolli_prototype_1998}
Ermolli, I., Fofi, M., Bernacchia, C., {et~al.} 1998, Solar Physics, 177, 1

\bibitem[{Ermolli {et~al.}(2009{\natexlab{a}})Ermolli, Marchei, Centrone,
  Criscuoli, Giorgi, \& Perna}]{ermolli_digitized_2009}
Ermolli, I., Marchei, E., Centrone, M., {et~al.} 2009{\natexlab{a}}, Astronomy
  and Astrophysics, 499, 627

\bibitem[{Ermolli {et~al.}(2013)Ermolli, Matthes, Dudok~de Wit, Krivova,
  Tourpali, Weber, Unruh, Gray, Langematz, Pilewskie, Rozanov, Schmutz,
  Shapiro, Solanki, \& Woods}]{ermolli_recent_2013}
Ermolli, I., Matthes, K., Dudok~de Wit, T., {et~al.} 2013, Atmospheric
  Chemistry \& Physics, 13, 3945

\bibitem[{Ermolli {et~al.}(2009{\natexlab{b}})Ermolli, Solanki, Tlatov,
  Krivova, Ulrich, \& Singh}]{ermolli_comparison_2009}
Ermolli, I., Solanki, S.~K., Tlatov, A.~G., {et~al.} 2009{\natexlab{b}}, The
  Astrophysical Journal, 698, 1000

\bibitem[{Fredga(1971)}]{fredga_comparison_1971}
Fredga, K. 1971, Solar Physics, 21, 60

\bibitem[{Fröhlich(2013)}]{frohlich_total_2013}
Fröhlich, C. 2013, Space Science Reviews, 176, 237

\bibitem[{Garcia {et~al.}(2011)Garcia, Sobotka, Klvana, \&
  Bumba}]{garcia_synoptic_2011}
Garcia, A., Sobotka, M., Klvana, M., \& Bumba, V. 2011, Contributions of the
  Astronomical Observatory Skalnate Pleso, 41, 69

\bibitem[{Gardner(1947)}]{gardner_validity_1947}
Gardner, I.~C. 1947, Validity of the cosine-fourth-power law of illumination
  (National Bureau of Standards)

\bibitem[{Giorgi {et~al.}(2005)Giorgi, Ermolli, Centrone, \&
  Marchei}]{giorgi_calibration_2005}
Giorgi, F., Ermolli, I., Centrone, M., \& Marchei, E. 2005, Memorie della
  Societa Astronomica Italiana, 76, 977

\bibitem[{Goldman \& Chen(2005)}]{goldman_vignette_2005}
Goldman, D. \& Chen, J.-H. 2005, in Tenth {IEEE} {International} {Conference}
  on {Computer} {Vision}, 2005. {ICCV} 2005, Vol.~1, 899--906 Vol. 1

\bibitem[{Haigh(2007)}]{haigh_sun_2007}
Haigh, J.~D. 2007, Living Reviews in Solar Physics, 4, 2

\bibitem[{Hanaoka(2013)}]{hanaoka_long-term_2013}
Hanaoka, Y. 2013, Journal of Physics Conference Series, 440, 2041

\bibitem[{Harvey \& White(1999)}]{harvey_magnetic_1999}
Harvey, K.~L. \& White, O.~R. 1999, The Astrophysical Journal, 515, 812

\bibitem[{Johannesson {et~al.}(1998)Johannesson, Marquette, \&
  Zirin}]{johannesson_10-year_1998}
Johannesson, A., Marquette, W.~H., \& Zirin, H. 1998, Solar Physics, 177, 265

\bibitem[{Kahil {et~al.}(2017)Kahil, Riethmüller, \&
  Solanki}]{kahil_brightness_2017}
Kahil, F., Riethmüller, T.~L., \& Solanki, S.~K. 2017, The Astrophysical
  Journal Supplement Series, 229, 12

\bibitem[{Kariyappa \& Pap(1996)}]{kariyappa_contribution_1996}
Kariyappa, R. \& Pap, J.~M. 1996, Solar Physics, 167, 115

\bibitem[{Kariyappa \& Sivaraman(1994)}]{kariyappa_variability_1994}
Kariyappa, R. \& Sivaraman, K.~R. 1994, Solar Physics, 152, 139

\bibitem[{Kopp(2016)}]{kopp_magnitudes_2016}
Kopp, G. 2016, Journal of Space Weather and Space Climate, 6, A30

\bibitem[{Krivova {et~al.}(2007)Krivova, Balmaceda, \&
  Solanki}]{krivova_reconstruction_2007}
Krivova, N.~A., Balmaceda, L., \& Solanki, S.~K. 2007, Astronomy and
  Astrophysics, 467, 335

\bibitem[{Krivova {et~al.}(2003)Krivova, Solanki, Fligge, \&
  Unruh}]{krivova_reconstruction_2003}
Krivova, N.~A., Solanki, S.~K., Fligge, M., \& Unruh, Y.~C. 2003, Astronomy and
  Astrophysics, 399, L1

\bibitem[{Krivova {et~al.}(2010)Krivova, Vieira, \&
  Solanki}]{krivova_reconstruction_2010}
Krivova, N.~A., Vieira, L. E.~A., \& Solanki, S.~K. 2010, Journal of
  Geophysical Research (Space Physics), 115, 12112

\bibitem[{Lean {et~al.}(1995)Lean, Beer, \& Bradley}]{lean_reconstruction_1995}
Lean, J., Beer, J., \& Bradley, R. 1995, Geophysical Research Letters, 22, 3195

\bibitem[{Lefebvre {et~al.}(2005)Lefebvre, Ulrich, Webster, Varadi, Javaraiah,
  Bertello, Werden, Boyden, \& Gilman}]{lefebvre_solar_2005}
Lefebvre, S., Ulrich, R.~K., Webster, L.~S., {et~al.} 2005, Memorie della
  Societa Astronomica Italiana, 76, 862

\bibitem[{Lites {et~al.}(2014)Lites, Centeno, \& McIntosh}]{lites_solar_2014}
Lites, B.~W., Centeno, R., \& McIntosh, S.~W. 2014, Publications of the
  Astronomical Society of Japan, 66, S4

\bibitem[{Livingston \& Sheeley(2008)}]{livingston_limits_2008}
Livingston, W. \& Sheeley, Jr., N.~R. 2008, The Astrophysical Journal, 672,
  1228

\bibitem[{Livingston \& Wallace(2003)}]{livingston_suns_2003}
Livingston, W. \& Wallace, L. 2003, Solar Physics, 212, 227

\bibitem[{Livingston {et~al.}(2007)Livingston, Wallace, White, \&
  Giampapa}]{livingston_sun-as--star_2007}
Livingston, W., Wallace, L., White, O.~R., \& Giampapa, M.~S. 2007, The
  Astrophysical Journal, 657, 1137

\bibitem[{Loukitcheva {et~al.}(2009)Loukitcheva, Solanki, \&
  White}]{loukitcheva_relationship_2009}
Loukitcheva, M., Solanki, S.~K., \& White, S.~M. 2009, Astronomy and
  Astrophysics, 497, 273

\bibitem[{Makarov {et~al.}(2004)Makarov, Tlatov, \&
  Callebaut}]{makarov_secular_2004}
Makarov, V.~I., Tlatov, A.~G., \& Callebaut, D.~K. 2004, Proceedings of the
  International Astronomical Union, 2004, 49

\bibitem[{Mein \& Ribes(1990)}]{mein_spectroheliograms_1990}
Mein, P. \& Ribes, E. 1990, Astronomy and Astrophysics, 227, 577

\bibitem[{Mickaelian {et~al.}(2007)Mickaelian, Nesci, Rossi, Weedman, Cirimele,
  Sargsyan, Erastova, Gigoyan, Mikayelyan, Massaro, Gaudenzi, Houck, Barry,
  D'Amante, \& Germano}]{mickaelian_digitized_2007}
Mickaelian, A.~M., Nesci, R., Rossi, C., {et~al.} 2007, Astronomy and
  Astrophysics, 464, 1177

\bibitem[{Nesme-Ribes {et~al.}(1996)Nesme-Ribes, Meunier, \&
  Collin}]{nesme-ribes_fractal_1996}
Nesme-Ribes, E., Meunier, N., \& Collin, B. 1996, Astronomy and Astrophysics,
  308, 213

\bibitem[{Peck \& Rast(2015)}]{peck_photometric_2015}
Peck, C.~L. \& Rast, M.~P. 2015, The Astrophysical Journal, 808, 192

\bibitem[{Pierce \& Slaughter(1977)}]{pierce_solar_1977}
Pierce, A.~K. \& Slaughter, C.~D. 1977, Solar Physics, 51, 25

\bibitem[{Priyal {et~al.}(2013)Priyal, Singh, Ravindra, Priya, \&
  Amareswari}]{priyal_long_2013}
Priyal, M., Singh, J., Ravindra, B., Priya, T.~G., \& Amareswari, K. 2013,
  Solar Physics

\bibitem[{Pötzi(2008)}]{potzi_scanning_2008}
Pötzi, W. 2008, Central European Astrophysical Bulletin, 32, 9

\bibitem[{Ravindra {et~al.}(2013)Ravindra, Priya, Amareswari, Priyal, Nazia, \&
  Banerjee}]{ravindra_digitized_2013}
Ravindra, B., Priya, T.~G., Amareswari, K., {et~al.} 2013, Astronomy and
  Astrophysics, 550, 19

\bibitem[{Schrijver {et~al.}(1989)Schrijver, Cote, Zwaan, \&
  Saar}]{schrijver_relations_1989}
Schrijver, C.~J., Cote, J., Zwaan, C., \& Saar, S.~H. 1989, The Astrophysical
  Journal, 337, 964

\bibitem[{Shapiro {et~al.}(2011)Shapiro, Schmutz, Rozanov, Schoell,
  Haberreiter, Shapiro, \& Nyeki}]{shapiro_new_2011}
Shapiro, A.~I., Schmutz, W., Rozanov, E., {et~al.} 2011, Astronomy and
  Astrophysics, 529, 67

\bibitem[{Singh {et~al.}(2012)Singh, Belur, Raju, Pichaimani, Priyal,
  Gopalan~Priya, \& Kotikalapudi}]{singh_determination_2012}
Singh, J., Belur, R., Raju, S., {et~al.} 2012, Research in Astronomy and
  Astrophysics, 12, 472

\bibitem[{Skumanich {et~al.}(1975)Skumanich, Smythe, \&
  Frazier}]{skumanich_statistical_1975}
Skumanich, A., Smythe, C., \& Frazier, E.~N. 1975, The Astrophysical Journal,
  200, 747

\bibitem[{Solanki \& Fligge(2000)}]{solanki_reconstruction_2000}
Solanki, S.~K. \& Fligge, M. 2000, Space Science Reviews, 94, 127

\bibitem[{Solanki {et~al.}(2013)Solanki, Krivova, \&
  Haigh}]{solanki_solar_2013}
Solanki, S.~K., Krivova, N.~A., \& Haigh, J.~D. 2013, Annual Review of
  Astronomy and Astrophysics, 51, 311

\bibitem[{Steinhilber {et~al.}(2009)Steinhilber, Beer, \&
  Fröhlich}]{steinhilber_total_2009}
Steinhilber, F., Beer, J., \& Fröhlich, C. 2009, Geophysical Research Letters,
  36, L19704

\bibitem[{Tlatov {et~al.}(2009)Tlatov, Pevtsov, \& Singh}]{tlatov_new_2009}
Tlatov, A.~G., Pevtsov, A.~A., \& Singh, J. 2009, Solar Physics, 255, 239

\bibitem[{Vieira {et~al.}(2011)Vieira, Solanki, Krivova, \&
  Usoskin}]{vieira_evolution_2011}
Vieira, L. E.~A., Solanki, S.~K., Krivova, N.~A., \& Usoskin, I. 2011,
  Astronomy and Astrophysics, 531, 6

\bibitem[{Walton {et~al.}(1998)Walton, Chapman, Cookson, Dobias, \&
  Preminger}]{walton_processing_1998}
Walton, S.~R., Chapman, G.~A., Cookson, A.~M., Dobias, J.~J., \& Preminger,
  D.~G. 1998, Solar Physics, 179, 31

\bibitem[{Wang {et~al.}(2005)Wang, Lean, \& Sheeley}]{wang_modeling_2005}
Wang, Y.-M., Lean, J.~L., \& Sheeley, Jr., N.~R. 2005, The Astrophysical
  Journal, 625, 522

\bibitem[{White \& Livingston(1978)}]{white_solar_1978}
White, O.~R. \& Livingston, W. 1978, The Astrophysical Journal, 226, 679

\bibitem[{White \& Livingston(1981)}]{white_solar_1981}
White, O.~R. \& Livingston, W.~C. 1981, The Astrophysical Journal, 249, 798

\bibitem[{Willis {et~al.}(2013)Willis, Coffey, Henwood, Erwin, Hoyt, Wild, \&
  Denig}]{willis_greenwich_2013}
Willis, D.~M., Coffey, H.~E., Henwood, R., {et~al.} 2013, Solar Physics, 288,
  117

\bibitem[{Worden {et~al.}(1998{\natexlab{a}})Worden, White, \&
  Woods}]{worden_evolution_1998}
Worden, J.~R., White, O.~R., \& Woods, T.~N. 1998{\natexlab{a}}, The
  Astrophysical Journal, 496, 998

\bibitem[{Worden {et~al.}(1998{\natexlab{b}})Worden, White, \&
  Woods}]{worden_plage_1998}
Worden, J.~R., White, O.~R., \& Woods, T.~N. 1998{\natexlab{b}}, Solar Physics,
  177, 255

\bibitem[{Yeo {et~al.}(2014)Yeo, Krivova, Solanki, \&
  Glassmeier}]{yeo_reconstruction_2014}
Yeo, K.~L., Krivova, N.~A., Solanki, S.~K., \& Glassmeier, K.~H. 2014,
  Astronomy and Astrophysics, 570, A85

\bibitem[{Zharkova {et~al.}(2003)Zharkova, Ipson, Zharkov, Benkhalil,
  Aboudarham, \& Bentley}]{zharkova_full-disk_2003}
Zharkova, V.~V., Ipson, S.~S., Zharkov, S.~I., {et~al.} 2003, Solar Physics,
  214, 89

\end{thebibliography}

\clearpage
\appendix
\section{Analysis of the synthetic data}
\label{sec:results_synth_ap}

In this section we will introduce each subset of synthetic data individually, providing information on their characteristics and how they were created. We will then show all the results for the errors we get when calculating the NSB and the CLV corrected calibrated images by applying our technique. The results obtained are also summarised in Table \ref{tab:syntheticdatasubsets}.

\subsection*{Subset 1: Linear relation without inhomogeneities}
We selected 500 Rome/PSPT images that sample the period 2000--2014 regularly, we computed the logarithm of each image and applied a linear CC to them to emulate the response of the plate. 
This set thus represents almost ideal density images without any large scale inhomogeneities or exposure problems, allowing to estimate the precision of the proposed method in the presence of AR.

\begin{figure}[h!]
	\centering
	\includegraphics[width=1\linewidth]{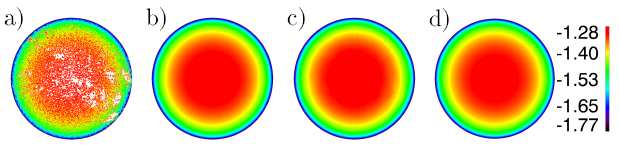}
	\caption{Example of application of our method to one image from subset 1. \textbf{a)} is the density image derived from the Rome/PSPT observation (saturated to the range of the QS for illustration purposes), \textbf{b)} is the imposed CLV, \textbf{c)} is the calculated NSB, and  \textbf{d)} is the calculated SB. The colour bar is the same for all 4 images.}
	\label{fig:psptexample}
\end{figure}

Figure \ref{fig:psptexample} shows an example of results obtained on one image from the subset 1, where \textbf{a)} is the density image derived from the Rome/PSPT observation, \textbf{b)} is the imposed CLV, \textbf{c)} is the calculated NSB and \textbf{d)} is the calculated SB. 

\begin{figure}
	\centering
	\includegraphics[width=1.0\linewidth]{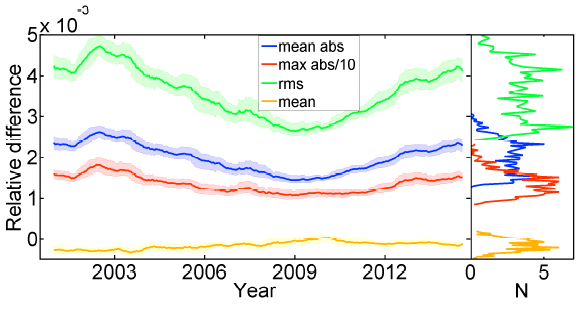}
	\caption{{\it Left:} Relative difference between the NSB and the imposed background for all images of subset 1 (disc positions within  0.98$R$). For clarity,  the maximum difference is  divided by 10. The solid lines are 30 point averages and the shaded surfaces denote the asymmetric 1$\sigma$ intervals. {\it Right:} Distribution functions of the difference values.}
	\label{fig:synth1_nsb}
\end{figure}

Figure \ref{fig:synth1_nsb} shows results derived by comparing the NSB retrieved from the proposed method and the imposed background for all the data of subset 1 (disc positions within  0.98$R$).  
Throughout this analysis the relative differences are given in absolute values (unsigned), but we also provide the signed average difference.
The red curve displays the maximum relative difference divided by 10 to improve the clarity of the plot.
The mean values of the maximum and average relative difference between the retrieved NSB (SB) and the CLV intensity  imposed on all images of subset 1 
are 0.013 (0.002) and  0.002 (0.0007), respectively; the RMS difference is 0.004 (0.001). On average only 0.5\% of the pixels show differences exceeding the average value $\pm 3 \sigma$ for the NSB. 
All above values show that the proposed method is able to retrieve radially symmetric backgrounds with great accuracy, even though this assumption is not made during the processing. 

\begin{figure}
	\centering
	\includegraphics[width=1.0\linewidth]{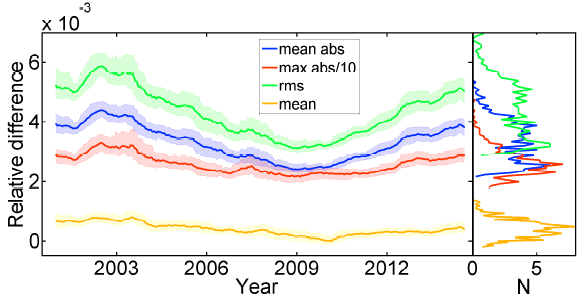}
	\caption{{\it Left}: Relative difference between the contrast image retrieved using our method and original undegraded data  for all images of subset 1 (disc positions within  0.98$R$). Labels are as in Fig. \ref{fig:synth1_nsb}. {\it Right}: Distribution functions of the difference values.}
	\label{fig:synth1_flat}
\end{figure}

Figure \ref{fig:synth1_flat} shows results obtained by comparing the contrast image after the calibration performed with the proposed method and the original data; the mean values of the maximum unsigned and average relative difference are  0.024 and 0.004, respectively,  and the RMS difference is 0.005.

The calculated slopes for the CC derived from the method are in good agreement with the imposed one. 
The results we get have a slope of on average 0.504 with standard deviation 0.002, compared to the imposed one of 0.5.

To estimate the accuracy of the method when applied on data unaffected by intensity variations between adjacent lines, we analysed subset 1 one more time but now bypassing \textit{step 2.5} of the background calculation (Sect. \ref{sec:backgroundcomputation}).
We found that the method's accuracy is on average a factor of 2 higher than obtained by applying the line fittings. In particular,  the mean values of the maximum unsigned and average relative difference between the retrieved NSB (contrast) and the CLV intensity (background) imposed on all images of subset 1 
are lower than 0.02 (0.013) and  0.0002 (0.0002), respectively; the RMS difference is 0.0027 (0.002).

\label{sec:results_synth1}
\subsection*{Subset 2: Different linear relations without inhomogeneities}

This subset was created to test the efficiency of the proposed method to account for different linear relations of the CC. 
Due to the changes in the photographic plates or films that were used, the different conditions of observations, development or digitization, the historical data have different CC. 
Therefore this subset tests the accuracy of our method to restore good quality images with different CC, but with no large-scale inhomogeneities or exposure problems. 
We selected 10 Rome/PSPT images covering the period 2000--2014 and imposed on them 20 different linear relations with slopes in the range 0.1--4.0.
Subset 2 therefore consists of 200 images.

\begin{figure}
	\centering
	\includegraphics[width=1\linewidth]{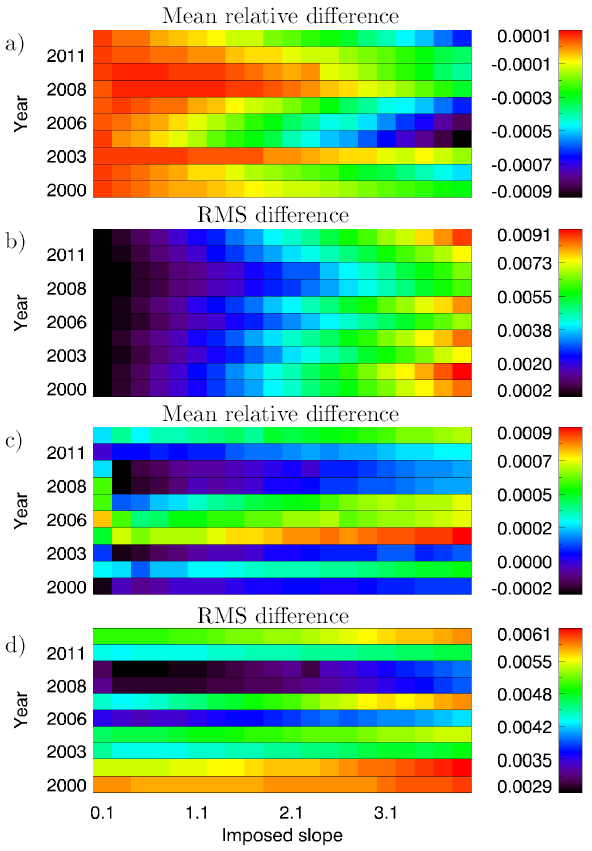}
	\caption{Colour-coded errors resulting from application of our method to images of subset 2. Each box of a given colour corresponds to a different observation (there are 200 boxes in all). Each row (column) of boxes shows results derived from a given observation (imposed CC relation). Note the different scale of the colour code, as represented by the corresponding colour bars. \textbf{a)} mean relative and \textbf{b)} RMS difference between the NSB and the imposed background for all images, \textbf{c)} mean relative and \textbf{d)} RMS difference between the retrieved contrast image and the original data. The values are given for disc positions within 0.98$R$.}
	\label{fig:synth2_errors_sb_max98}
\end{figure}

Figure \ref{fig:synth2_errors_sb_max98} shows the relative differences obtained by comparing images retrieved using our method and original undegraded data; panels \textbf{a)} and \textbf{b)} (\textbf{c)} and \textbf{d)}) refer to values of NSB (contrast) images. Each box represents a different image of subset 2,  the rows (columns) show values derived from synthetic images created with different Rome/PSPT contrast observations (different imposed CC relations). The values on the $y$-axis represent the year when the used contrast Rome/PSPT observations were taken. The different colours give the errors according to the colour bar next to the plot. 
The relative differences between the retrieved  NSB (SB) and imposed background increase with increasing slope of the relation applied, remaining however relatively small. The maximum unsigned relative difference measured is 0.03 (0.01) for the extreme case, but on average remains $<0.0009$ and maximum of RMS difference is 0.009 (0.004). 

The relative difference between the contrast image derived from our method and the original data reaches up to 0.07 for individual pixels in the worst case (on average $<0.0009$); the  RMS difference is less than 0.008.

\label{sec:results_synth2}
\subsection*{Subset 3: Non-linear relations without inhomogeneities}
\label{sec:data_pspt2}
\begin{figure}
	\centering
	\includegraphics[width=1\linewidth]{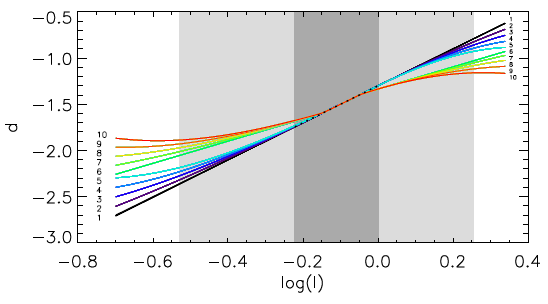}
	\caption{CC relations imposed on the data of subset 3.  Note that the parts of each curve for $\log{I}<-0.1$ (corresponding to under-exposure) can be combined with the part of the same or any other curve for $\log{I}\ge-0.1$ (over-exposure). E.g. 10,1 implies very strong under-exposure, no over-exposure, while 10,10 corresponds to very strong under- and overexposure. 
	The dark (light) grey shaded areas denote the range of values of the QS (entire Rome/PSPT image taken on 21/08/2000 which was used to create the synthetic images shown in Fig. \ref{fig:synthexamples1} -- \ref{fig:synthexamples2}). The dotted black line denotes the CC imposed on data of subset 1. Find details in Section \ref{sec:testonsyntheticdata}.}
	\label{fig:psptnonlinear}
\end{figure}

This subset was created to test the effects of applying a linear calibration on data that were constructed by imposing non-linear CC relations.
We used the same 10 images as employed for subset 2, computed the logarithm of each image and applied a series of non-linear relations on them. 
This way we can evaluate the accuracy of the proposed method for data suffering over- and under-exposure and to quantify the expected deviations introduced when applying a linear calibration on them. 
Figure  \ref{fig:psptnonlinear} shows the imposed relations. We used 2 separate functions to describe under-exposure and over-exposure. For each exposure problem we used 10 different strength levels and considered all combinations of these functions (in total 100 cases per image). 
The first 5 cases of over- and under-exposure (labelled 1-5 in Fig. \ref{fig:psptnonlinear}) produce images with intensity values belonging to the QS lying on the linear part of the CC, while for all the rest of the cases only a small part of the QS lies on the linear part of the CC.
Subset 3 consists of 1000 images.

\begin{figure}
	\centering
	\includegraphics[width=1\linewidth]{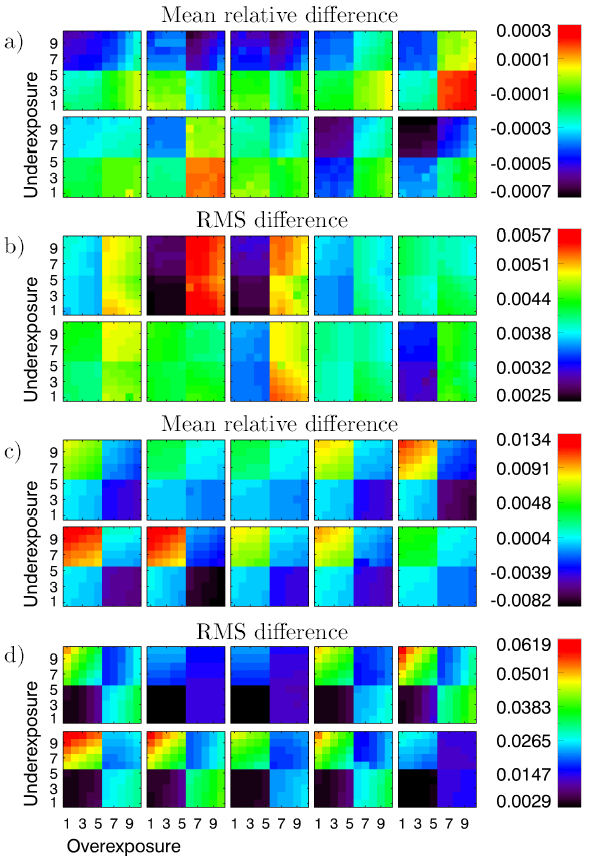}
	\caption{Colour-coded errors from application of our method to images of subset 3. The colour of each box (so-to-say each pixel of the plotted pattern) represents the error introduced when processing one particular Rome/PSPT image with different introduced exposure problems. Each square composed of $10\times 10$ boxes in a given panel corresponds to synthetic images created from a given Rome/PSPT image, with the rows (columns) within the square showing the errors for different combinations of under- (over-) exposures considered (the employed CC curves are shown in Fig. \ref{fig:psptnonlinear}). The colour bar applies to all boxes in a given panel. \textbf{a)} mean relative and \textbf{b)} RMS difference between the NSB and the imposed background for all images, \textbf{c)} mean relative and \textbf{d)} RMS difference between the retrieved contrast image and the original data. The values are given for disc positions within 0.98$R$.}
	\label{fig:synth3_errors_flat_max98}
\end{figure}

Figure \ref{fig:synth3_errors_flat_max98} shows the relative differences between the images derived with our  method and the original undegraded data of subset 3; panels \textbf{a)} and \textbf{b)} (\textbf{c)} and \textbf{d)}) refer to values of NSB (contrast) images. Each $10\times10$ box is for synthetic images derived from different Rome/PSPT contrast images. In each panel, rows (columns) show values derived from different levels of over-exposure (under-exposure).

Overall the accuracy introduced by the method decreases with increasing exposure problems. Nonetheless the errors remain relatively small (maximum unsigned relative difference is 0.02 for the extreme case and 0.006 maximum unsigned RMS difference) and are always comparable with the results from Subset 1. The inaccuracy increases  in the cases when over-exposure affects the QS. 
The differences in SB remain extremely low for no under-exposure problems affecting the QS, maximum unsigned relative difference is 0.007, while maximum unsigned RMS difference is 0.005.

The  maximum unsigned relative difference obtained by comparing the retrieved and original undegraded contrast images is usually $>0.1$ and reaches up to 0.95 for the extreme exposure problems (case 10 in Fig. \ref{fig:psptnonlinear}); RMS is $<0.62$.
These errors affect mostly the bright features. The maximumunsigned  differences remain $<0.2$ for the maximum over-exposure and under-exposure considered which does not affect the QS (cases 1--5  in Fig. \ref{fig:psptnonlinear}) and decrease with decreasing exposure problems. 
The errors are usually reasonable for the cases when there is no over-exposure on the QS. 

\begin{figure}
	\centering
	\includegraphics[width=1\linewidth]{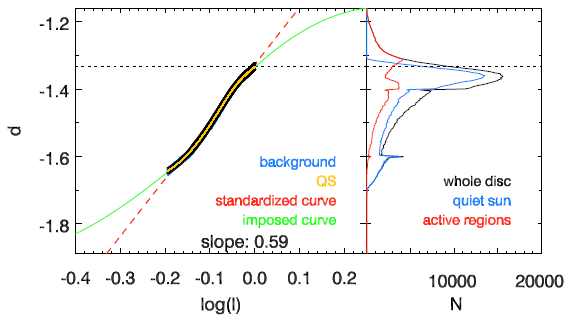}
	\caption{\textit{Left}: Imposed CC (green), standardized CC derived with our method (red), CC  estimated from  the QS (orange) with its 1$\sigma$ level (black circles), and the whole background (blue, lying almost perfectly behind the black circles) of the synthetic image of subset 3 derived from Rome/PSPT observation taken on 21/08/2000 (shown in Fig. \ref{fig:synthexamples1} \textbf{S3/1)}), which corresponds to an extreme case of over- and under-exposure on the QS considered in our study (level 10 for both under- and over- exposure). Shown also is the slope of the derived CC. \textit{Right}: Distribution of density values for the QS (blue), the AR (red) and the whole disc (black). The dashed line in both panels denotes the highest value of the QS CLV.}
	\label{fig:20000821K12ffc_clvrem_nsb4_15_d3rddegree_04_10rd478yrpierceyr}
\end{figure}

Figure  \ref{fig:20000821K12ffc_clvrem_nsb4_15_d3rddegree_04_10rd478yrpierceyr} shows the CC constructed for the extreme exposure problems considered. 
The green, blue, and red  curves display  the imposed relation, the measured QS CLV, and the standardized linear relation, respectively.
For all the data, our method allows calculating the NSB accurately enough so that the obtained CLV reproduces  the imposed relation for almost the whole range of QS values. 
In particular, the maximum unsigned relative difference between the imposed CC and the one recovered by the method is less than 0.005. The accuracy drops slightly for the last point at the limb and only for the cases of severe under-exposure, rising to 0.017.

\label{sec:results_synth3}
\subsection*{Subset 4: Different disc sizes with linear CC without inhomogeneities in the image}
The photographic plates from different observatories have different sizes. Even among the images of the same archive there are variations due to changes of the spectroheliograph and of the photographic plates. 
The digitization affects the number of pixels covered by the disc as well. The disc size is also affected by the seasonal variation of the Sun--Earth distance.
Therefore, this subset was created to test the efficiency of the proposed method when applied to datasets that have a different disc size. 
We selected 10 Rome/PSPT images covering the period 2000--2014 and re-sampled them, so that each was represented in the form of solar images with 10 different radii between 100 and 550 pixels.
The range of radii was defined to include the sampling present in most available datasets, however the upper limit was dictated by the number of pixels covered by the disc on Rome/PSPT observations. 
Subset 4 consists of 100 images.

\begin{figure}
	\centering
	\includegraphics[width=1\linewidth]{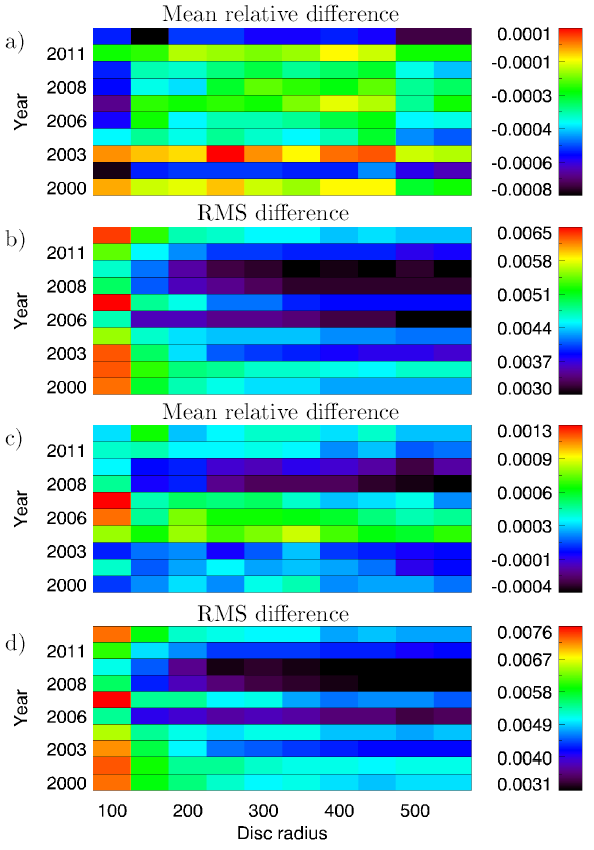}
	\caption{Colour-coded relative differences between the images retrieved by applying our method to images of subset 4 and the underlying original, undegraded Rome/PSPT images. Each box (with a given colour) corresponds to a different synthetic image; row (column) shows results derived from a given Rome/PSPT observation (imposed radius). Note the different scale of the colour code, as represented by the corresponding colour bars. \textbf{a)} mean relative and \textbf{b)} RMS difference between the NSB and the imposed background within a given subfigure, \textbf{c)} mean relative and \textbf{d)} RMS difference between the retrieved contrast image and the original data. The values are given for disc positions within 0.98$R$.}
	\label{fig:synth4_errors_sb_max98}
\end{figure}

Figure \ref{fig:synth4_errors_sb_max98} shows the relative differences between the images derived with our method and the original Rome/PSPT data; panels \textbf{a)} and \textbf{b)} (\textbf{c)} and \textbf{d)}) refer to relative differences between NSB (contrast) images, each (coloured) box for a different image of subset 4. 
In each panel,  rows (columns) show values derived from different observations (imposed radii).
The relative  difference between the retrieved  NSB (SB) and imposed background decreases with increasing radius of the disc; the maximum unsigned difference measured is 0.03 (0.007) for the images with the smallest radius,  average difference is $<0.0008$, and maximum of RMS difference is 0.004 (0.001).

The relative difference between the contrast image derived with our method and the original data reaches up to 0.056; the average differences are $<0.007$; the RMS differences are $<0.009$. 

\label{sec:results_synth4}
\subsection*{Subset 5: Vignetting with linear CC without inhomogeneities}
This subset was created to evaluate the effects of vignetting on the performance of the proposed method. 
We took the same 10 Rome/PSPT observations as employed above and imposed on them the same radially symmetric CLV as in subset 1, along with the vignetting function $c \times \cos^4{s}$ \citep{gardner_validity_1947,goldman_vignette_2005} in intensity, where $s$ is the distance of each pixel to the centre of the disc and $c$ is a constant that takes 10 discrete values between 0 and 0.5. 
This vignetting function causes maximum decrease in normalised intensities between 0 near the disc centre and 0.3 (which corresponds to 50\%) near the limb in the density images.	
Subset 5 consists of 100 images.

\begin{figure}
	\centering
	\includegraphics[width=1\linewidth]{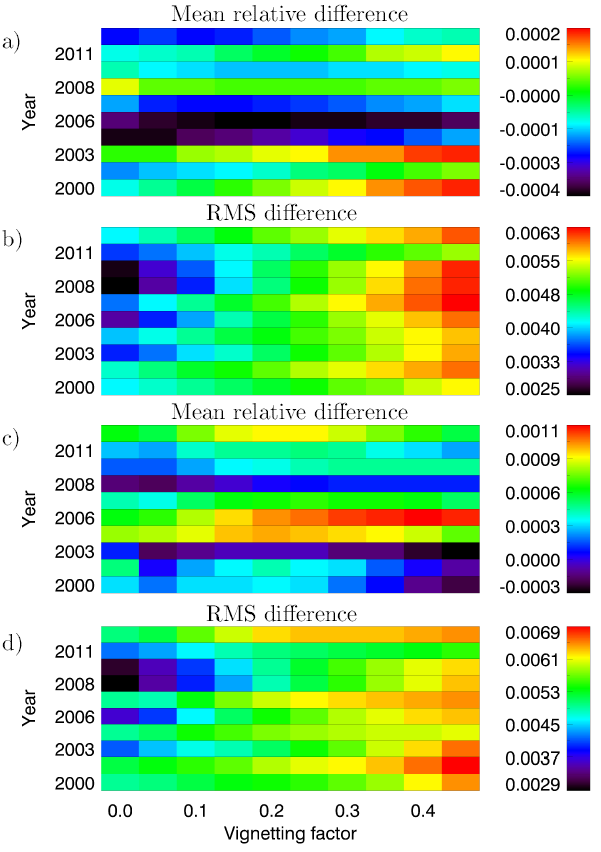}
	\caption{Colour-coded relative differences between the images retrieved by applying our method to images of subset 5 and the underlying original, undegraded Rome/PSPT images. Each box corresponds to a different synthetic image. Each row (column) shows results derived from a given Rome/PSPT observation (imposed vignetting). Note the different scale of the colour code, as represented by the corresponding colour bars. \textbf{a)} mean relative and \textbf{b)} RMS difference between the NSB and the imposed background for all images, \textbf{c)} mean relative and \textbf{d)} RMS difference between the retrieved contrast image and the original Rome/PSPT data. The values are given for disc positions within 0.98$R$.}
	\label{fig:synth5_errors_nsb_max98}
\end{figure}

Figure \ref{fig:synth5_errors_nsb_max98} shows the relative differences between the images derived with our method and the original data. Panels \textbf{a)} and \textbf{b)} (\textbf{c)} and \textbf{d)}) refer to values of NSB (contrast) images, each box corresponds to a different image of subset 5. 
In each panel,  rows (columns) show values derived from different observations (imposed vignetting).
The relative  difference between the retrieved  NSB (SB) and imposed background increases with increasing strength of vignetting (value of $c$); the maximum unsigned difference measured is 0.018 (0.014) for the extreme case, the average differences are $<0.0004$, and maximum of RMS difference is 0.005 (0.016). 

The relative difference between the contrast image derived with our method and the original data reaches up to 0.065; the average differences are $<0.0003$; the RMS differences are $<0.008$.

\begin{figure}
	\centering
	\includegraphics[width=1\linewidth]{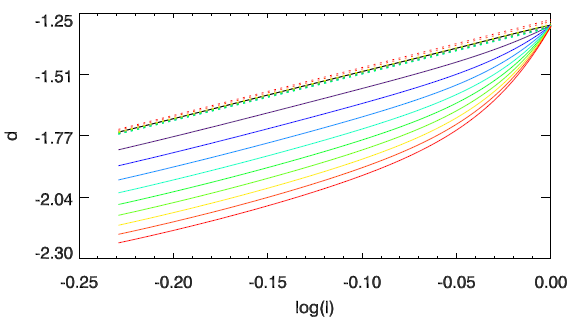}
	\caption{CC for synthetic images with vignetting (subset 5) created from Rome/PSPT observation taken on 21/08/2000. The colours denote the magnitude of the vignetting, with red for the strongest case and black for no vignetting. The solid curves are derived with the values measured on the density image, while the dotted curves are after the correction for the vignetting by our method.}
	\label{fig:synth5_characteristiccurves}
\end{figure}

Figure \ref{fig:synth5_characteristiccurves} shows the effect of the vignetting on the CC. 
As intended, vignetting decreases non-linearly the density towards the limb. This results in significantly wrong slopes for the computed CC if the vignetting is not accounted for. Calibrating the data with the wrong slope of the CC results in lower contrasts than the ones from the undegraded data and the contrast tends to decrease with increasing magnitude of vignetting.

\label{sec:results_synth5}
\subsection*{Subset 6: Varying magnitude of inhomogeneities with linear CC}
\label{sec:data_synth3}
\begin{figure}[h!]
	\centering
		\includegraphics[width=1\linewidth]{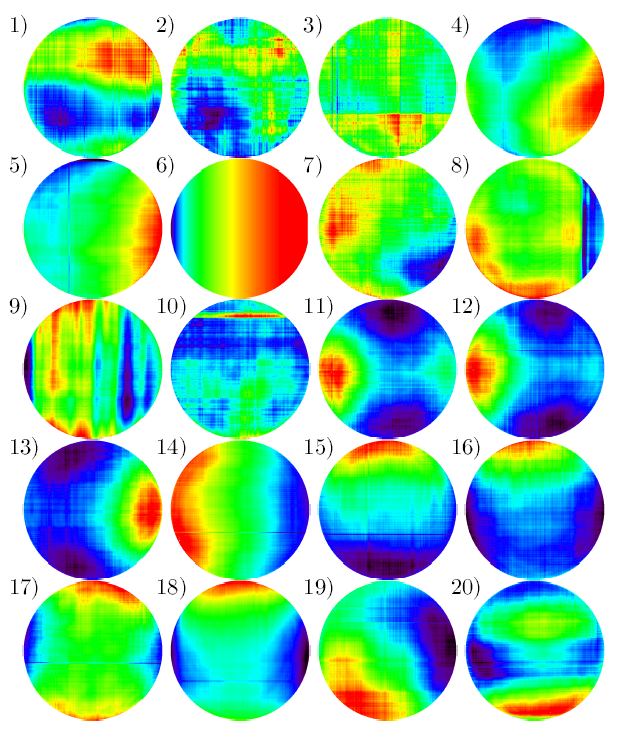}
	\caption{Inhomogeneities used to create images of subsets 6 and 8. The first 5 backgrounds were derived from Ar observations, the sixth one is artificial and introduces a gradient of a fourth root function over the disc, 7--10 from Ko, 11--18 from MW, and 19--20 from Mi images.}
	\label{fig:synth5_background}
	
\end{figure}

This subset was created to evaluate the effects of the varying levels of inhomogeneities on the results derived with the proposed method. 
The magnitude of the inhomogeneities is compared to the range of values of the CLV and we construct for each background 10 cases with inhomogeneities whose amplitude lies in the range [0.1 -- 1.0] of the CLV. 
The backgrounds used here are shown in Fig. \ref{fig:synth5_background}.
The facular patterns overlain on these backgrounds were derived from 10 Rome/PSPT observations in the period 2000--2014. 
Subset 6 consists of 2000 images, resulting from 20 backgrounds, 10 amplitudes for each background, each applied to the 10 chosen Rome/PSPT images.

\begin{figure}
	\centering
		\includegraphics[width=1\linewidth]{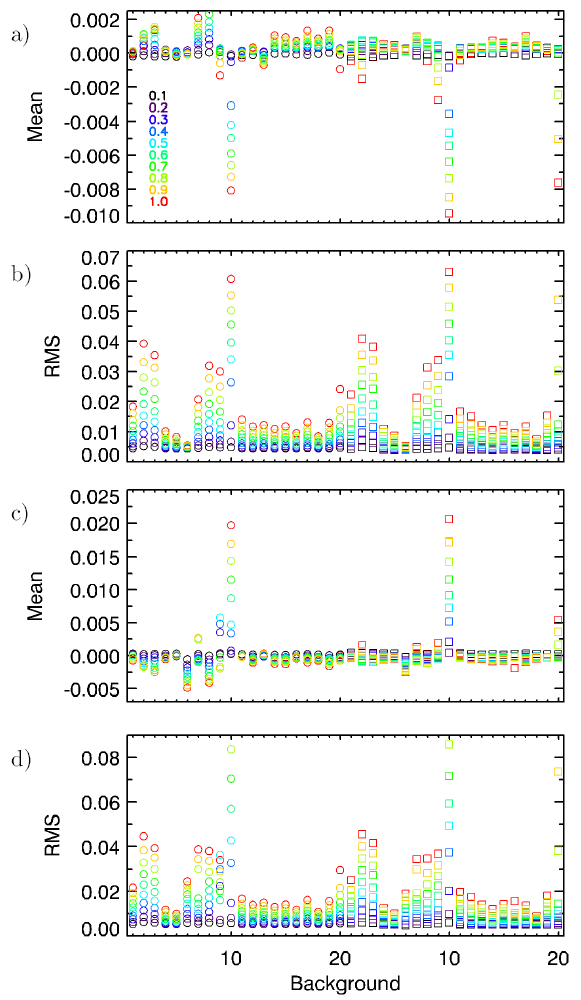}
	\caption{Errors in the images retrieved with our method applied to subset 6. The circles (squares) correspond to errors in images based on Rome/PSPT data taken at a high (low) activity period. The colours denote the different magnitudes of the inhomogeneities. The scale (relative to the background intensity variation from the limb to disc centre) is given in the topmost frame. The various backgrounds are numbered following Fig. \ref{fig:synth5_background}. \textbf{a)} mean relative and \textbf{b)} RMS difference between the NSB and the imposed background for all images, \textbf{c)} mean relative and \textbf{d)} RMS difference between the retrieved contrast image and the original data. The values are given for disc positions within 0.98$R$.}
	\label{fig:synth6_nsb}
\end{figure}

The relative differences of the NSB derived with our method and the original data of subset 6 can be seen in Fig. \ref{fig:synth6_nsb}. 
We show the results for only two Rome/PSPT images, one taken at high and one at low solar activity periods, in order to make it easier to discern the differences between the results obtained from the various inhomogeneities and levels.
In particular, in Fig. \ref{fig:synth6_nsb} the x axis represents the different backgrounds, the circles show results for the high activity image, while the squares are for the low activity image and the colours signify the varying levels of inhomogeneities (black implying low amplitudes, while red is for large amplitudes).

Unsurprisingly, the errors introduced when retrieving the intensity image using the proposed method increase with magnitude of the inhomogeneities on the analysed image. 
In many images the errors remain relatively low, e.g. background No. 6 in Fig. \ref{fig:synth5_background}, where the maximum difference is 0.007. 
The maximum unsigned difference is obtained for the images with very bright/dark small artefacts, while larger-scale inhomogeneities and small artefacts with mild brightness are reliably processed by our method.
The errors are highest when density variations in adjacent image lines are significantly brighter than their surroundings, with a maximum relative difference of 0.5. 
The maximum differences for NSB remain $<0.1$ for most of the backgrounds even in the extreme cases that were considered.

The results for the rest of the data in subset 6 are very similar to the ones presented here. We did not find any significant dependence of the results on the level of solar activity.

\label{sec:results_synth6}
\subsection*{Subset 7: Different CLV with linear CC}
\begin{figure}
	\centering
	\includegraphics[width=1.0\linewidth]{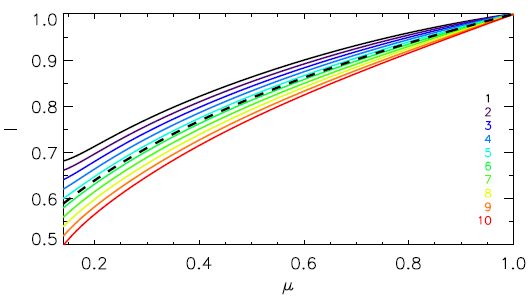}
	\caption{Intensity CLV imposed on the images of subset 7. The dashed black line denotes the average CLV measured on Rome/PSPT observations. The colours from black to red correspond to the cases 1--10 shown in Fig. \ref{fig:synth7_errors_nsb_max98}.}
	\label{fig:difclv}
\end{figure}

Subset 7 was created to study the effects of applying our method on data that have a different CLV than the one we use to standardize them. 
The data could have a different CLV because the observation was centred at a different wavelength than the core of the Ca~II~K line, or possibly due to a different bandwidth (e.g. due to a different spectral resolution or slit width). 
We used again 10 Rome/PSPT images to derive the facular pattern and on each one of those imposed 10 different functions for the quiet Sun (seen in Fig. \ref{fig:difclv}), describing deviations from the one we measure in modern Rome/PSPT observations.
Subset 7 consists of 100 images.

\begin{figure}
	\centering
	\includegraphics[width=1\linewidth]{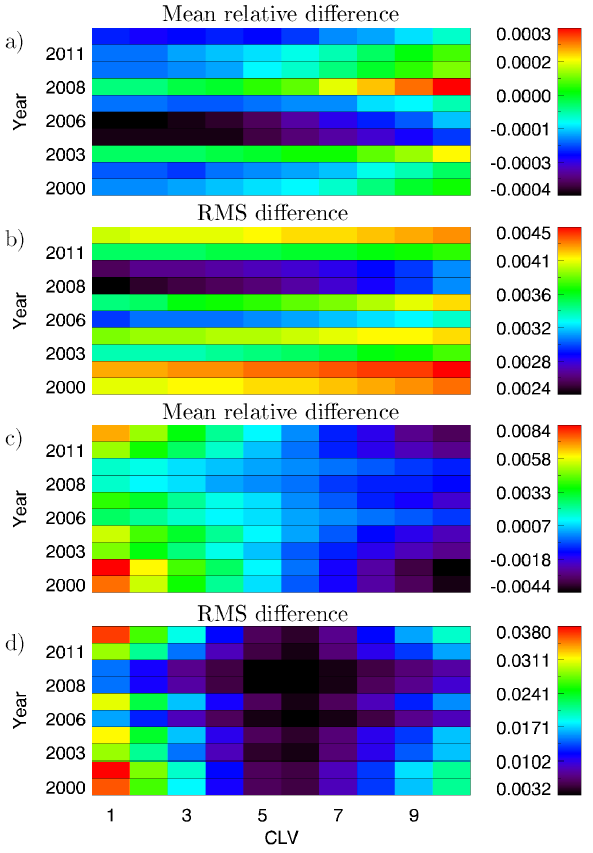}
	\caption{Colour-coded relative differences between the images retrieved by applying our method to images of subset 7 and the underlying original, undegraded Rome/PSPT images. Each box corresponds to a different observation. Each row (column) of boxes shows results derived from a given observation identified by the year in which it was recorded (imposed CLV, numbered as in Fig. \ref{fig:difclv}). Note the different scale of the colour code, as represented by the corresponding colour bars. \textbf{a)} mean relative and \textbf{b)} RMS difference between the NSB and the imposed background for all images, \textbf{c)} mean relative and \textbf{d)} RMS difference between the retrieved contrast image and the original data. The values are given for disc positions within 0.98$R$.}
	\label{fig:synth7_errors_nsb_max98}
\end{figure}

Figure \ref{fig:synth7_errors_nsb_max98} shows the relative differences between the corrected and calibrated images derived with our method and the original data; panels \textbf{a)} and \textbf{b)} (\textbf{c)} and \textbf{d)}) refer to values of NSB (contrast) images, each box for a different image of subset 7. 
In each panel,  rows (columns) show values derived from different observations (imposed CLV).
The relative  difference between the retrieved  NSB (SB) and imposed background has a maximum value of 0.019 (0.007), the average differences are $<0.0004$, and the maximum RMS difference is 0.004 (0.003).

We found that the different CLV affect more the calculation of the slope of the CC. The errors for the contrast images increase with the RMS differences getting up to 0.038, while the maximum differences remain $<10$\% for 3 CLV surrounding the average curve we use. 

\label{sec:results_synth7_diffclv}
\subsection*{Subset 8: Random problems}
Subset 8 was created to study the performance of our method on data where every image suffers from a different random set of problems.
We used 2000 Rome/PSPT images to derive the facular pattern and on each one of those imposed a random function for the quiet Sun (shown in Fig. \ref{fig:difclv}), describing deviations from the one we measure in modern Rome/PSPT observations. We also added a random pattern of inhomogeneities that were used for subset 6 (Fig. \ref{fig:synth5_background}), with a random level of strength for each image. We convert the images to density by applying a CC in the form of a 3rd degree polynomial with randomly selected parameters within the following ranges: [-4.6, -1.3] for the constant term, [0.5, 4.0] for the linear term, [-0.03, 0.00] for the quadratic term, and [-0.03, 0.00] for the cubic term. A random level of vignetting was also added.
Subset 8 consists of 2000 images.

\begin{figure}
\centering
\includegraphics[width=1\linewidth]{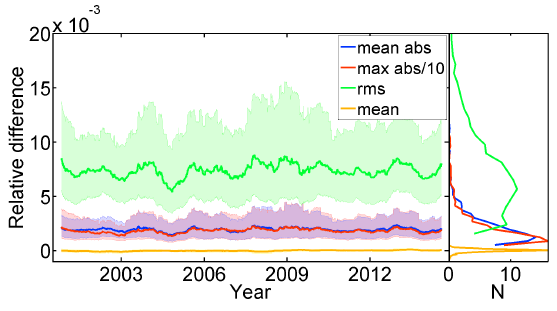}
\caption{{\it Left}: Relative difference between the NSB and the imposed background (within $0.98R$) for all the synthetic data of subset 8. RMS difference (green), mean absolute difference (blue), mean difference (orange) and maximum difference (red). These differences are plotted vs. the date on which the original Rome/PSPT images (that were randomly distorted) were recorded. Note that the maximum difference values have been divided by 10 to plot them on the same scale as the other quantities. The solid lines are 100 point averages and the shaded surfaces denote the asymmetric 1$\sigma$ interval.  {\it Right}: Distribution of the relative difference values.}
\label{fig:synth8_nsb}
\end{figure}

The results for this subset are presented in the main text in Sec. \ref{sec:overallsynthetic}, here in Fig. \ref{fig:synth8_nsb} we also show the plot with the relative errors for the NSB calculation.

\end{document}